\documentclass[%
 reprint,
unsortedaddress,
 amsmath,amssymb,
 aps,
]{revtex4-2}

\usepackage{graphicx}
\usepackage{dcolumn}
\usepackage{bm}

\begin{document}


\title{Local Analysis of Heterogeneous Intracellular Transport: Slow and Fast moving Endosomes}

\author{Nickolay Korabel}
\affiliation{Department of Mathematics, The University of Manchester, M13 9PL, UK}
\email{nickolay.korabel@manchester.co.uk}

\author{Daniel Han}
\affiliation{
Department of Mathematics, The University of Manchester, M13 9PL, UK}%
\affiliation{
School of Biological Sciences, The University of Manchester, M13 9PT, UK}%
\affiliation{
Biological Physics, Department of Physics and Astronomy, The University of Manchester, M13 9PL, UK}%

\author{Alessandro Taloni}
\affiliation{
CNR - Consiglio Nazionale delle Ricerche, Istituto dei Sistemi Complessi, via dei Taurini 19, 00185 Roma, Italy
}%

\author{Gianni Pagnini}
\affiliation{BCAM - Basque Center for Applied Mathematics, Mazarredo 14, E-48009 Bilbao, Basque Country - Spain
}%
\affiliation{Ikerbasque - Basque Foundation for Science, Plaza Euskadi 5, E-48009 Bilbao, Basque Country - Spain
}%

\author{Sergei Fedotov}
 \affiliation{Department of Mathematics, The University of Manchester, M13 9PL, UK}

\author{Viki Allan}
\affiliation{%
School of Biological Sciences, The University of Manchester, M13 9PT, UK
}%

\author{Thomas Andrew Waigh}
\affiliation{%
Biological Physics, Department of Physics and Astronomy, The University of Manchester, M13 9PL, UK}%
\email{t.a.waigh@manchester.ac.uk}

\date{\today}

\begin{abstract}
Trajectories of endosomes inside living eukaryotic cells are highly heterogeneous in space and time and diffuse anomalously due to a combination of viscoelasticity, caging, aggregation and active transport. Some of the trajectories display switching between persistent and anti-persistent motion while others jiggle around in one position for the whole measurement time. By splitting the ensemble of endosome trajectories into slow moving sub-diffusive and fast moving super-diffusive endosomes, we analyzed them separately. The mean squared displacements and velocity auto-correlation functions confirm the effectiveness of the splitting methods. Applying the local analysis, we show that both ensembles are characterized by a spectrum of local anomalous exponents and local generalized diffusion coefficients. Slow and fast endsomes have exponential distributions of local anomalous exponents and power law distributions of generalized diffusion coefficients. This suggests that heterogeneous fractional Brownian motion is an appropriate model for both fast and slow moving endosomes. This article is part of a Special Issue entitled: "Recent Advances In Single-Particle Tracking: Experiment and Analysis" edited by Janusz Szwabiński and Aleksander Weron.
\end{abstract}

\maketitle


\section{\label{Intro}Introduction}

Intracellular transport of organelles, such as endosomes, has been described by anomalous diffusion caused by different mechanisms \cite{Klages,HF}. Various models have been proposed to describe it, such as Fractional Brownian motion (FBM), continuous time random walks and fractional Langevin equations \cite{Metzler2014}. However, which of these models is the best is a current topic of much debate.

To decipher which mechanism is at work and determine the appropriate mathematical model to describe it, a large ensemble of trajectories is necessary. Modern experimental techniques facilitate the tracking of large ensembles of intracellular objects for considerable amounts of time. Therefore, the extraction of meaningful statistical information from trajectories is becoming an important issue. The traditional statistical analysis of trajectories includes quantification of ensemble evolution in time and space using the ensemble-averaged mean squared displacements (EMSD), time-averaged MSD (TMSD), probability density functions of displacements and correlation functions. As the accessible measurement time in experiments increases with better live-cell microscopy techniques, the accurate analysis of single trajectories has become possible \cite{BGM}. New methods of trajectory analysis were developed, such as local time-averaged MSD \cite{Heinrich}, first passage probability analysis \cite{WaighBOOK} and time-averaged diffusion coefficients \cite{Akimoto}. 

Improved microscopy imaging, tracking and analysis methods revealed the intrinsic spatial and temporal heterogeneity within individual trajectories of numerous biological processes \cite{Heinrich,ELife,Saxton,Granick1,Granick2,Lampo,Weiss2020,BaLysosomes,ManzoPRX,SadoonWang}. 
Significant progress has been made also in analysis and interpretation of superresolution single particle  trajectories \cite{Calderon, Holcman, Andersson, Weron, Weiss2014}.
Recently, individual trajectories of quantum dots in the cytoplasm of living cultured cells were found to perform sub-diffusive motion of the fractional Brownian FBM type with switching between two distinct mobility states \cite{Janczura}. In contrast to homogeneous systems, heterogeneous trajectories are most prominently described by broad distributions of diffusivities and anomalous exponents, an exponential probability distribution of diffusivities and a Laplace probability distribution of displacements \cite{Metzler2020}. These observations led to the development of various heterogeneous diffusion models \cite{Beck1,Molina,Mackala,Chertvy2014,Spakowitz,Chubynsky,Sposini,Chechkin,Wang,itto-beck-2021,KB2010,FK2015,FH2019,Sandev}.

Recently, the intracellular transport of endosomes in eukaryotic cells was shown to be described by spatio-temporal heterogeneous fractional Brownian motion (hFBM) with non-constant Hurst exponents \cite{N}. By analysing the local motion of endosomes, we found that it is characterized by power-law probability distributions of displacements and displacement increments, exponential probability distributions of local anomalous exponents and power-law probability distributions of local generalized diffusion coefficients. In this paper, we split the ensemble of endosomes into slow and fast moving vesicles 
which is 
the main difference between this study and Ref. \cite{N}. This splitting allows us to study sub-ensembles separately in addition to studying of the ATP driven active transport of endosomes. 
In particular, there is the central question: What is the appropriate mathematical model to describe the subdiffusive transport of slow moving endosomes? By analysing locally the slow and fast endosomal trajectories, we find that both are characterised by exponential distributions of anomalous exponents and power-law distributions of generalized diffusion coefficients. This suggests that hFBM is an appropriate model for both slow and fast endosomes.

Endosome trajectories are composed of segments of active and passive motion and therefore they could be further decomposed into directed runs and random motion. We segmented endosomal trajectories in this way in Ref. \cite{ELife}. In this manuscript, we separated endosome trajectories into superdiffusive trajectories and subdiffusive trajectories for their whole duration. Subdiffusive trajectories do not contain segments of directed movement and cannot be segmented further into active and passive motion. In contrast, fast superdiffusive trajectories can be further segmented. We leave the segmentation of fast trajectories into directed runs and random motion for future work.

\section{Materials and Methods}

\subsection{Experimental trajectories} 

We studied a large ensemble of two dimensional experimental trajectories, $\mathbf{r}(t)=\{x(t),y(t)\}$, of early endosomes in a stable MRC5 cell line expressing GFP-Rab5. Trajectories were obtained from tracking wide-field fluorescence microscopy videos (see \cite{ELife} for experimental details). 
We studied 103,361 experimental trajectories of early endosomes, the same data which was acquired in \cite{ELife}. An example of experimental trajectories is shown in Fig. \ref{figS0}. The endosomes were tracked using an automated tracking software (AITracker, based on a convolutional neural network)\cite{Newby}. Currently it is not yet feasible to determine the diameter of endosomes in these experiments, because they are diffraction limited. Thus, it was possible to track the centres of endosomes with sub-pixel accuracy, but not the sizes of the smaller endosomes (less than 200 nm). The duration of all trajectories, $T$, has a good fit to a power law distribution, $T^{-1.85}$ \cite{N}, which is a manifestation of the heterogeneity of the trajectories. Slow moving endosomes stay longer within the observation volume and therefore have longer trajectories than fast moving endosomes, leading to the emergence of the power-law probability distribution for the trajectoriesâ€™ duration.

\subsection{Splitting of ensemble into slow and fast moving endosomes} 

We split ensemble of trajectories into slow and fast moving endosomes using the distance traveled by endosomes:
\begin{equation}
R(t)=\sqrt{(x(t)-x(0))^2+(y(t)-y(0))^2}.
\end{equation}
Trajectories which possess active motion have periods of rapid increase or decrease of $R$ (Figure~\ref{fig1}a). Fast trajectories which have active motion are defined as $\mbox{max}\{R(t)\}>\epsilon$  and slow trajectories which exhibit only passive motion are defined by $\mbox{max}\{R(t)\}<\epsilon$. Here $\mbox{max}\{R(t)\}$ denotes the maximum values of $R(t)$ attained in the time interval $(0, t)$ and $\epsilon$ is the threshold. We choose the threshold $\epsilon=0.25 \; \mu$m. In the Appendix we show that changing the threshold to $\epsilon=0.2 \; \mu$m 
in the splitting does not qualitatively change the results. 
Therefore, we define fast moving endosomes as those that in the time interval $(0, t)$, experienced at least one period of active motion and the maximum distance travelled from the origin exceeds the threshold of $\epsilon=0.25$  $\mu$m. Otherwise, an endosome is defined as slow moving. Small variations of the threshold value do not affect the EMSDs of slow and fast moving endosomes, which suggests that the splitting method is robust (Figure~\ref{figS1B}). 

Changing the splitting threshold from $\mbox{max}\{R(t)\}=0.25$ $\mu$m to $\mbox{max}\{R(t)\}=0.2$ $\mu$m, the increase of the number of slow trajectories was $12\%$. Therefore, in addition to the method of splitting trajectories which uses the minimum travelled distance, we have also tested a second method, which makes use of the time-dependent Hurst exponent $H(t)$ neural network (NN) estimate  at the single trajectory level \cite{ELife}. The procedure is as follows: 1) estimate the time-dependent anomalous exponent $\alpha^{NN}$ using the NN; 2) if the anomalous exponent $\alpha^{NN}$ is super-diffusive $\alpha^{NN}(t)>1$ for more than $4$ consecutive time points, the endosome is considered as fast moving. Otherwise, the endosome is labelled as slow moving (see Figure~\ref{figS1}). The correct implementation of the NN procedure requires a minimum time window \cite{ELife} that is larger than the duration of some of the endosomal trajectories. Hence, short trajectories were discarded in this analysis. 
The similarity of the distributions of generalized diffusion coefficients (Figures A3B and A4B) suggests that the chosen threshold $\mbox{max}\{R(t)\}=0.25$ $\mu$m was reasonable.
Alternative methods of binary classification could be performed using the first passage probability analysis \cite{Rogers} or implementing the normalized radius of gyration of each trajectory \cite{He}. 

\subsection{Ensemble and time averaged mean squared displacements} 

From the two dimensional experimental trajectories $\mathbf{r}(t)=\{x(t),y(t)\}$, we calculated the ensemble-averaged mean squared displacement (EMSD) as
\begin{equation}
\mbox{EMSD}(t) = \frac{\left< \mathbf{r} \right>^2(t)}{l^2},
\label{emsd}
\end{equation}
where $l$ is the length scale which we choose $l=1$ $\mu$m,
\begin{equation}
\left< \mathbf{r} \right>^2(t) = \left< (x_i(t)-x_i(0))^2 + (y_i(t)-y_i(0))^2 \right>,
\label{emsd1}
\end{equation}
where the angular brackets denotes averaging over an ensemble of trajectories, $\left< A \right>=\sum_{i=1}^{N} A_i/N$ and $N$ is the number of trajectories in the ensemble. 

By fitting the EMSD to power law functions, the anomalous exponent $\alpha$ and the generalized diffusion coefficient $D_{\alpha}$ can be extracted using
\begin{equation}
\mbox{EMSD}(t) = 4 D_{\alpha} \left( \frac{t}{\tau} \right)^{\alpha},
\label{emsdfit}
\end{equation}
where $\alpha$ and $D_{\alpha}$ are constants which characterize averaged transport properties of ensemble of endosomal trajectories. The time scale $\tau=1$ sec and the length scale $l=1$ $\mu$m are introduced in order to make the generalized diffusion coefficient $D_{\alpha}$ dimensionless. 

The time-averaged mean squared displacement (TMSD) of an individual trajectory $\{x_i, y_i \}$ of a duration $T$ is calculated as:
\begin{equation}
\mbox{TMSD}_i(t) = \frac{\overline{\delta^2(t)}}{l^2}, 
\label{tmsd}
\end{equation}
where $l$ is the length scale which we choose $l=1$ $\mu$m and
\begin{equation}
\overline{\delta^2(t)} = \frac{ \int_{0}^{T-t} \left( x_i(t'+t)-x_i(t'))^2 + (y_i(t'+t)-y_i(t'))^2 \right) dt'}{T-t}.
\label{tmsd1}
\end{equation}
TMSDs of individual trajectories are averaged further over the ensemble of trajectories to get the ensemble-time-averaged MSD (E-TMSD): 
\begin{equation}
\mbox{E-TMSD}(t) = \left< \mbox{TMSD}_i(t) \right>,
\label{etmsd}
\end{equation}
where the angular brackets denotes averaging over an ensemble of trajectories as before.

\subsection{Local analysis of endosomal trajectories} 

The time-local statistical analysis has been implemented as follows. We considered only the portion of a single endosomal trajectory within a window of size $W$ and centered around the time $t$, i.e. $(t-W/2,t+W/2)$. We have calculated the TMSD within this chunk of trajectory only: this is the reason of the acronym L-TMSD, i.e. the local TMSD. As the experimental detection of the endosomal motion is achieved with the frame rate $1/\Delta t$ s$^{-1}$, hence $t=i \Delta t$ (here $i=0, 1, 2, ...$ is the time index) and $W=N\Delta t$, with $N>10$. The first $10$ points of the L-TMSD were fitted with the power-law function 
\begin{equation}
\mbox{L-TMSD} = 4 D^L(t) \left( \frac{t'}{\tau} \right)^{\alpha^L(t)},
\label{ltmsdfit}
\end{equation}
where $t'=10\Delta t$. $\alpha^L(t)$ and $D_{\alpha^L}(t)$ are  the local anomalous exponent and generalized diffusion coefficient respectively. We iterate this procedure by shifting the time window of a single $\Delta t$ ($i\to i+1$) till the end of the experimental endosomal trace, thus obtaining  $\alpha^L(t)$ and  $D_{\alpha^L}(t)$ along the entire trajectory. Notice that $\alpha^L$ and $D^L$ are not constants in time but they vary, being local properties of each endosomal trajectory. 

\subsection{The time and ensemble-time averaged velocity auto-correlation functions}

The time averaged auto-correlation function (TVACF) along a single trajectory  is defined as: 
\begin{equation}
\mbox{TVACF}_i(t) = \frac{ \int_{0}^{T-t -\tau} \vec{v}(t'+t) \vec{v}(t') dt'}{T-t - \tau},
\label{tvacf}
\end{equation}
where $\vec{v}=\frac{\vec{r}(t+\tau)-\vec{r}(t)}{\tau}$.
TVACFs of individual trajectories are averaged further over the ensemble of trajectories to get the ensemble-time averaged VACF (E-TVACF): 
\begin{equation}
\mbox{E-TVACF}(t) = \left< \mbox{TVACF}_i(t) \right>,
\label{etvacf}
\end{equation}
where the angular brackets denotes averaging over an ensemble of trajectories. The velocity autocorrelation function was suggested as a tool to distinguish between subdiffusion models \cite{Weber2010}. 

\section{Results}

\begin{figure}[htb]
\includegraphics[scale=0.3]{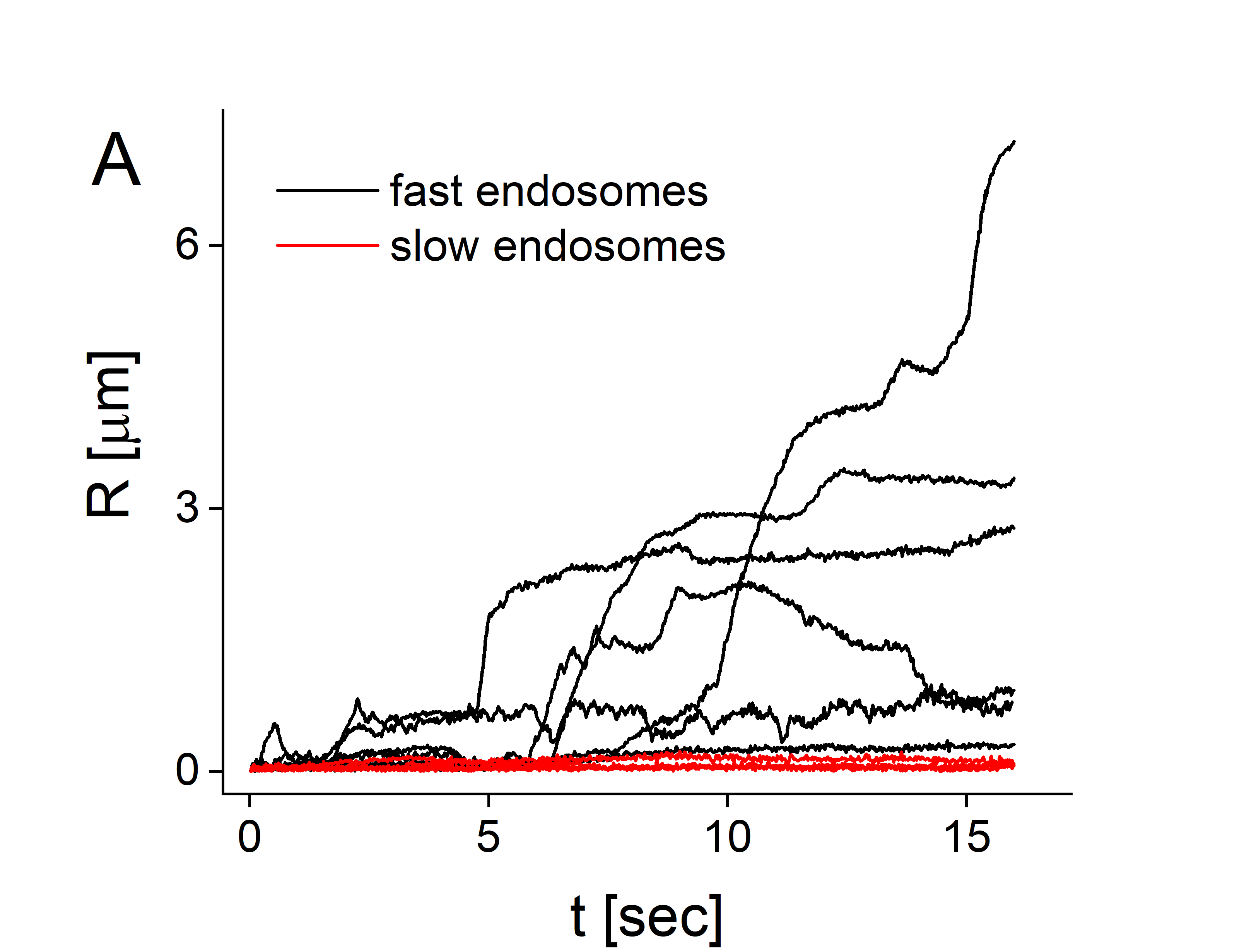}
\includegraphics[scale=0.3]{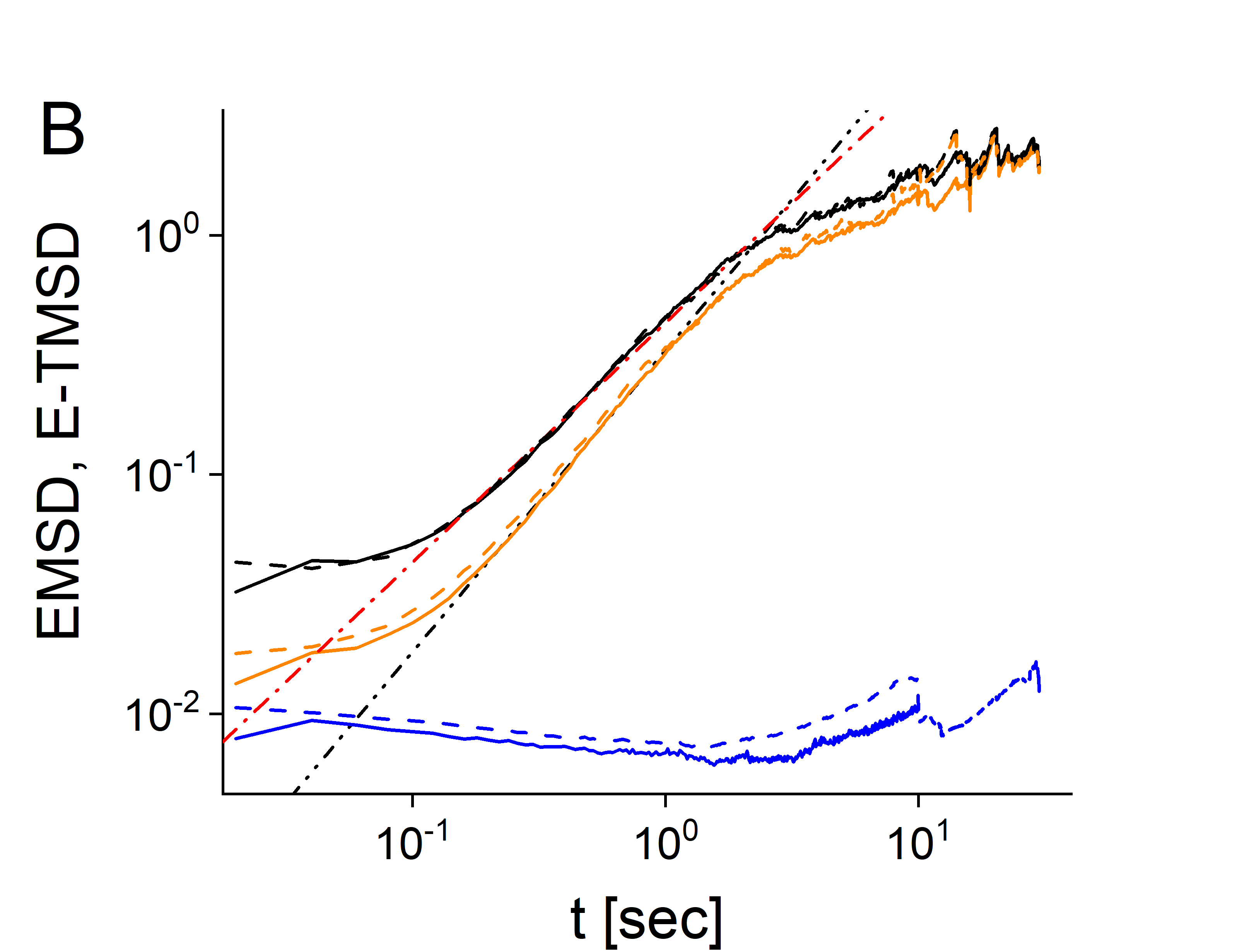}
\caption{Endosomes are split into slow and fast moving: (\textbf{A}) Distance $R(t)$ traveled by fast (black curves) and slow (red curves) endosomes ($9$ sample experimental trajectories are shown). Most experimental trajectories possess active motion visible as a rapid increase or decrease of $R$. (\textbf{B})  EMSDs (solid curves) and E-TMSDs (dashed curves) of fast (black curves) and slow (blue curves) endosomes compared with EMSD and E-TMSD of all trajectories (orange curves). The dashed-dotted and dashed-double-dotted lines represent $t^{1.26}$ and $t$ functions.
\label{fig1}}
\end{figure}

We split the ensemble of endosomes into slow and fast moving vesicles using the two methods described above (see Methods, Figure~\ref{fig1}A and Figure~\ref{figS1}). For both slow and fast endosomes, the EMSDs and the E-TMSDs show similar behaviour which suggest ergodicity (see Methods and Figure~\ref{fig1}B). MSDs of slow endosomes are not increasing in time, which confirms that these trajectories have no active periods of motion. 
Surprisingly, we found that both EMSDs and E-TMSDs of slow endosomes  are decreasing functions of time which to our knowledge was never observed before. We explain this behaviour in terms of the coupling between the average diffusivities of slow trajectories and their duration (see Figure \ref{figS1A} and the discussion below).
Conversely, MSDs of fast endosomes are increasing functions of time in the intermediate time scale $(0.2, 2)$ s. The anomalous exponent extracted from EMSD or E-TMSD of fast endosomes is $\alpha \simeq 1$, smaller than the anomalous exponent obtained by considering  all trajectories without distinction into fast or slow, i.e. $\alpha \simeq 1.26$. Notice that two sub-diffusive regimes characterize the MSD time behavior for fast and all trajectories. The first, at small time scales ($t\leq 10^{-1}$ s) can be  attributed to the measurement errors \cite{Weber,WaighReview,Doyle}; the second, at longer time scales ($t> 10$ s) was shown to be spurious and originate from the coupling of the trajectories' duration and their diffusivities \cite{N,Etoc}.  We suggest that, due to this coupling, the anomalous exponents deduced from the power-law fit of EMSD and E-TMSD, do not capture the essential characteristics of the endosome superdiffusive motility, nor shed light on its fundamental aspects. Therefore, in order to 
reveal the effect of the duration of trajectories on  the statistical analysis, we  
consider only trajectories longer than a certain threshold $T$ \cite{N}.

\begin{figure}[htb]
\includegraphics[scale=0.3]{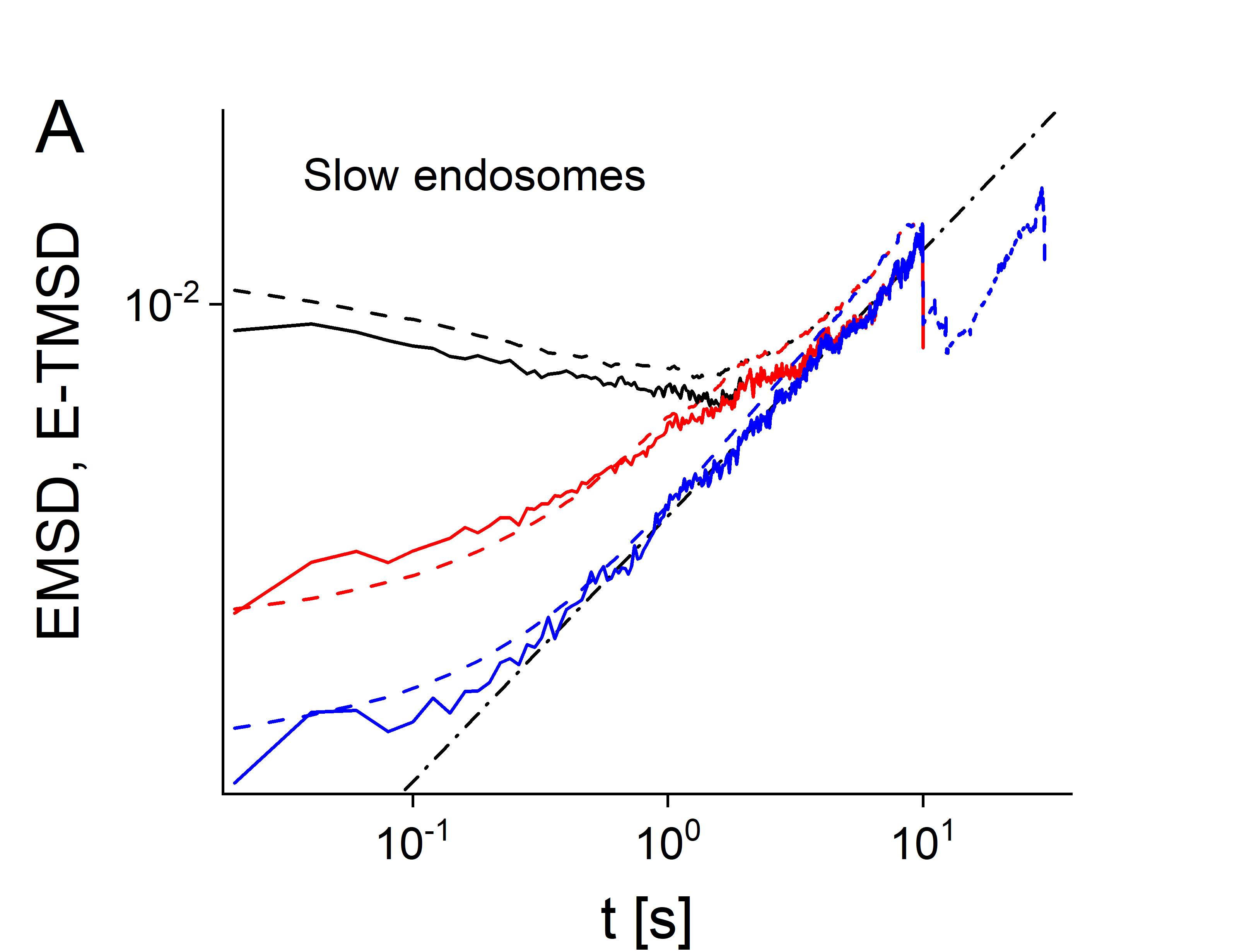}
\includegraphics[scale=0.3]{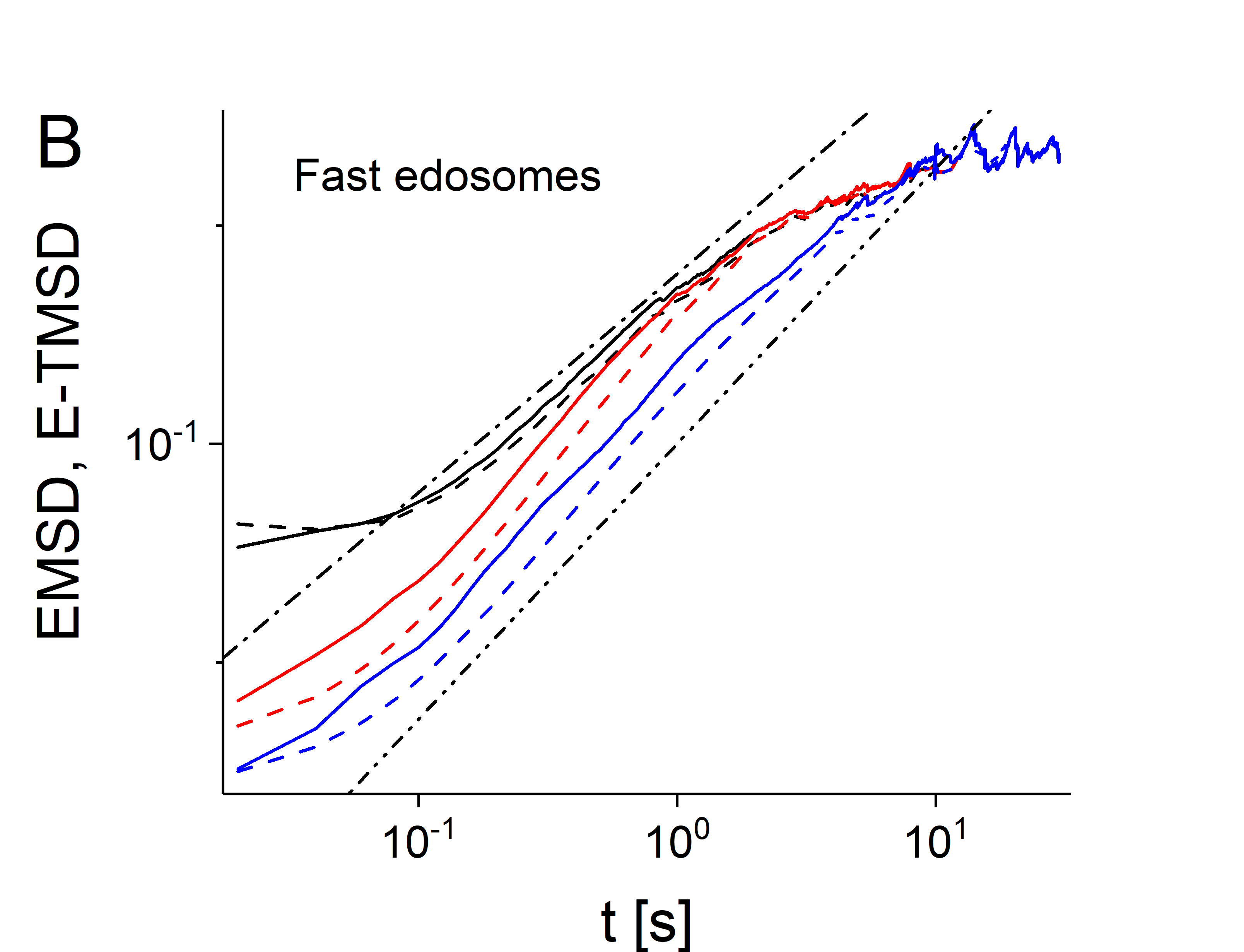}
\caption{EMSDs and E-TMSDs (solid and dashed curves) of experimental trajectories of slow (\textbf{A}) and fast moving endosomes (\textbf{B}). Black curves correspond to $T\rightarrow\infty$ s. Red and blue curves represent EMSDs and E-TMSDs of experimental trajectories which have duration longer than $2$ and $8$ seconds, respectively. The dashed-dotted line in (\textbf{A}) represents the function $t^{0.5}$. In (\textbf{B}), the dashed-dotted line and the dashed-double-dotted line represents the linear, $t$, and super-linear, $t^{1.26}$, functions respectively. 
\label{fig1A}}
\end{figure}

Figure~\ref{fig1A}A and B shows the EMSDs and E-TMSDs of slow and fast endosomes, considering only experimental trajectories with duration longer than $T$ seconds (2 or 8 sec). Unlike the slow moving endosomes, the  MSDs of  fast vesicles (Figure~\ref{fig1A}B) present similar qualitative behaviors  either by choosing $T=2$ s, $T=8$ s or no $T$ at all (all the fast molecules considered as in Fig.~\ref{fig1}B, black curve). However, in the intermediate regime, the superdiffusive behaviour becomes more and more apparent, $\propto t^{1.26}$, and stable.  In Ref. \cite{N} we have found that this process is described by the space-time heterogeneous FBM with the Hurst exponent $H$ that randomly switches between persistent $H>0.5$ and anti-persistent regimes $H<0.5$, together with the coupling between the diffusivity and duration of trajectories which account for spurious sub-diffusion at longer time scales.  Moreover the EMSD curves obtained for $T=2$ s and $T=8$ s deviates considerably  from the corresponding E-TMSD curves.
The MSDs of slow endosomes (Figure~\ref{fig1A}A) display very different, but ergodic, behaviour. For $0.01 < t < 2$ s, the MSDs of all slow endosomes decreases in time. On the other side, the MSDs of the sub-ensembles of slow endosomes with  $T=2$ s and $T=8$ s, reveal  sub-diffusive trends with $\alpha \sim 0.5$. As in the case of fast moving endosomes, we argue that this behaviour is due to the coupling between the diffusivity and duration of trajectories. Therefore we attempt to confirm this hypothesis, by performing simulations of an ensemble of heterogeneous FBM trajectories with constant Hurst exponent $H=0.25$ (see Figure~\ref{figS1A} below).  

\begin{figure}[htb]
\includegraphics[scale=0.3]{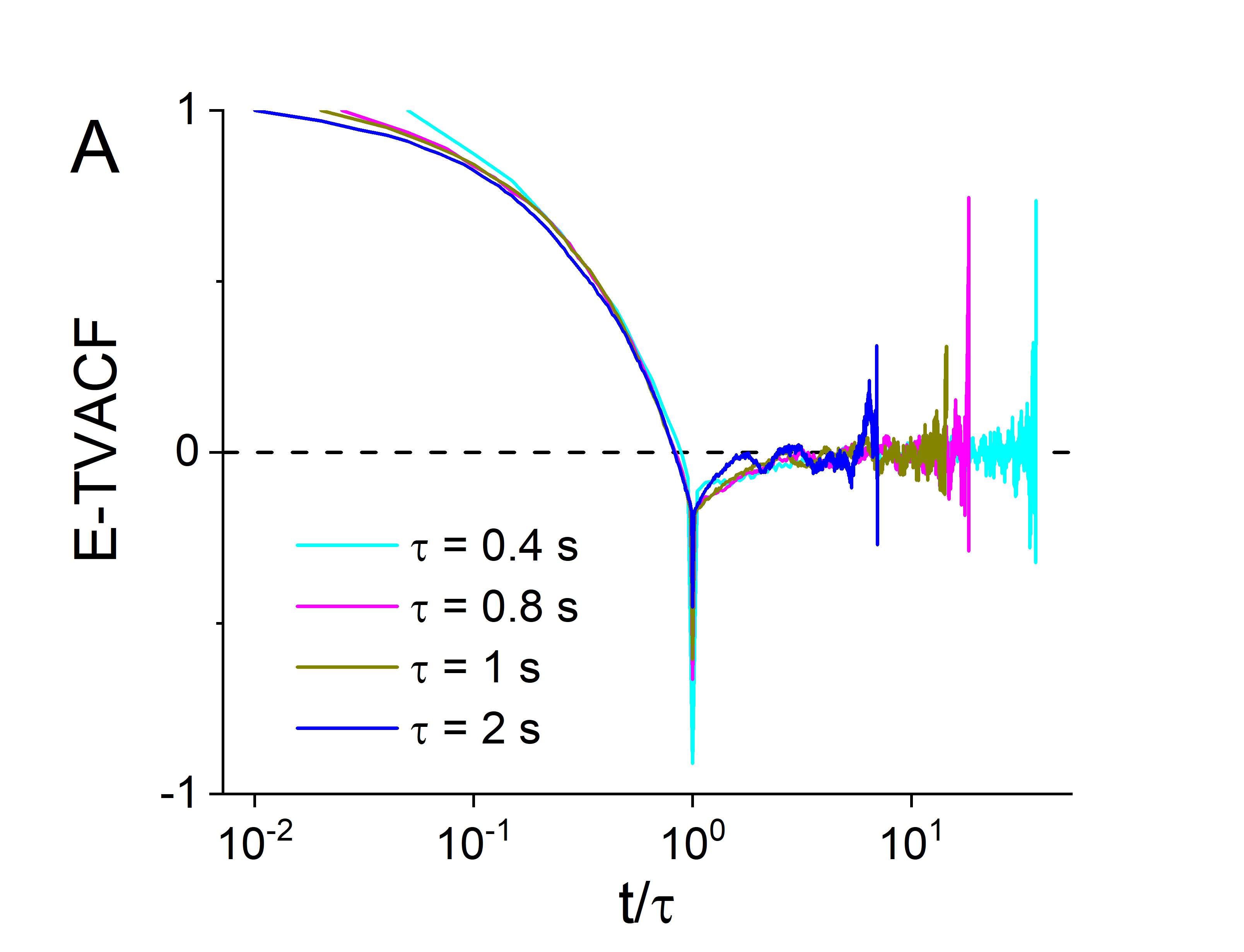}
\includegraphics[scale=0.3]{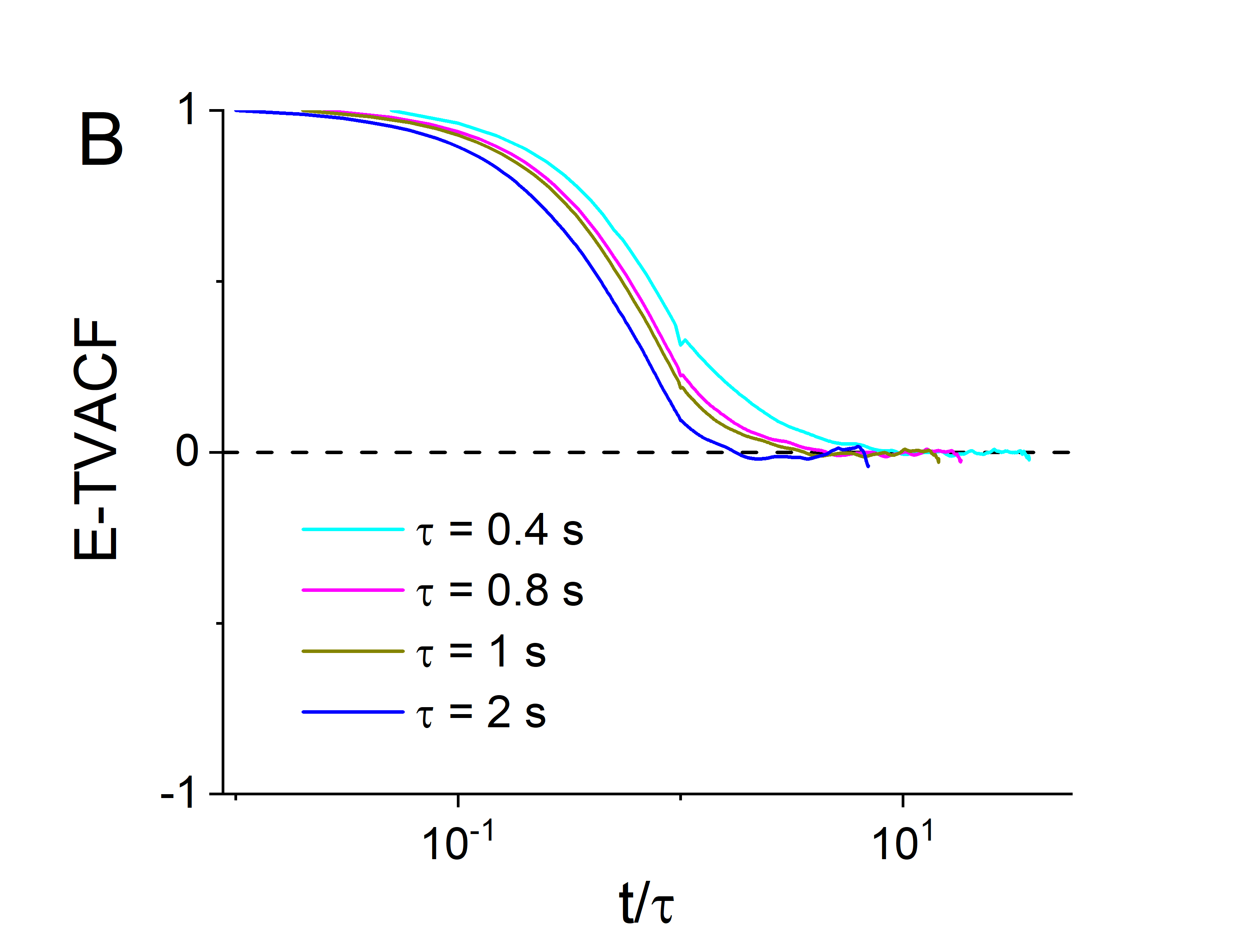}
\caption{Time-ensemble averaged VACF (E-TVACF) of experimental trajectories of slow (\textbf{A}) and fast endosomes (\textbf{B}) calculated for different $\tau$ given in the legend. 
\label{fig3}}
\end{figure}

\begin{figure}[htb]
\includegraphics[scale=0.3]{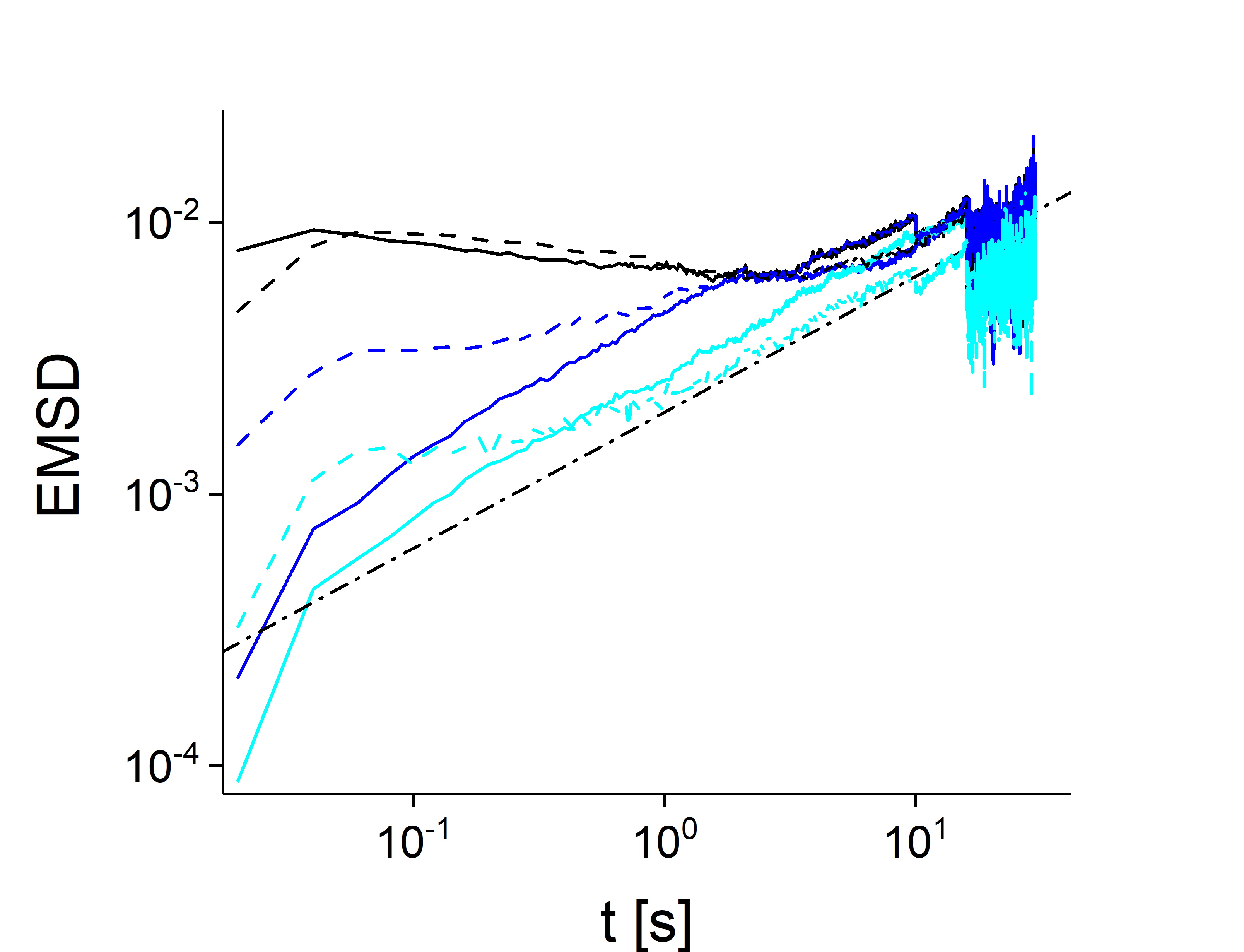}
\caption{EMSDs calculated for simulated hFBM trajectories (solid lines) as a function of time interval. Black curves correspond to EMSDs of all trajectories of slow endosomes and hFBM, blue curves are EMSDs of trajectories longer than $T=2$ s and cyan curves are EMSDs of trajectories longer than $T=8$ s. 
The sub-diffusive behaviour with the anomalous exponent $\alpha = 0.5$ is shown as the dashed-dotted line. The EMSDs of slow experimental endosomal trajectories are shown for comparison (dashed lines). Notice that hFBM trajectories were simulated without external noise (measurement error) which lead to discrepancy between simulated and experimental EMSDs at small time scale.
\label{figS1A}}
\end{figure}

The velocity auto-correlation functions (VACF) also confirm the effectiveness of this simple threshold splitting (Figure~\ref{fig3}A and B). Indeed, slow and fast endosomes have very different VACFs. Ensemble-time averaged VACFs (E-TVACFs) of fast endosomes (Figure~\ref{fig3}B) are positive as expected for super-diffusive motion. In contrast, E-TVACFs of slow endosomes have negative dips at $t=\tau$ and approach zero from negative values (Figure~\ref{fig3}A). Such behaviour is characteristic of FBM and the generalized Langevin equation, but cannot be reproduced by the CTRW model \cite{Metzler2014}.

To verify that heterogeneous FBM describes slow moving endosomes, we simulated an ensemble of hFBM trajectories. Individual hFBM trajectories were simulated with constant Hurst exponent $H = 0.25$. For standard FBM this would correspond to sub-diffusive MSDs, $\left< \mathbf{r}^2(t) \right> \sim t^{2H} \sim t^{0.5}$. The duration of hFBM trajectories was drawn from  the power-law distribution $\phi(T) \sim T^{-1.85}$, in accordance with  the experimental evidence \cite{N}. The generalized diffusion coefficients were chosen  inversely proportional to the duration of trajectories, i.e. $D \sim T^{-0.6}$. 
As shown in  Figure~\ref{figS1A}, the EMSDs of hFBM trajectories agree well with the experimental data. 

We now implement the local analysis \cite{N} to better characterize the slow and fast endosomal dynamics. We calculate the  local TMSDs (L-TMSD) for each experimental trajectory at various times $t$ ( Methods). From  the fit of L-TMSD to Eq.\ref{ltmsdfit} we extracted the local anomalous exponents $\alpha^L(t)$ and the local generalized diffusion coefficients $D_{\alpha^L}(t)$  for slow and fast endosomes separately.  
The local anomalous exponents $\alpha^L(t)$ and the local generalized diffusion coefficients $D_{\alpha^L}(t)$ appear to be positively correlated both for slow and fast endosomes (see Fig. \ref{figS5}). The origin of these correlations is not known and will be investigated in future publications. In Ref. \cite{N}, we found that PDFs of local anomalous exponents and local generalized diffusion coefficients  do not depend on the window size, nor on the time $t$ (stationary) and are best fitted with exponential and power law functions respectively. 
\begin{figure}[htb]
\includegraphics[scale=0.3]{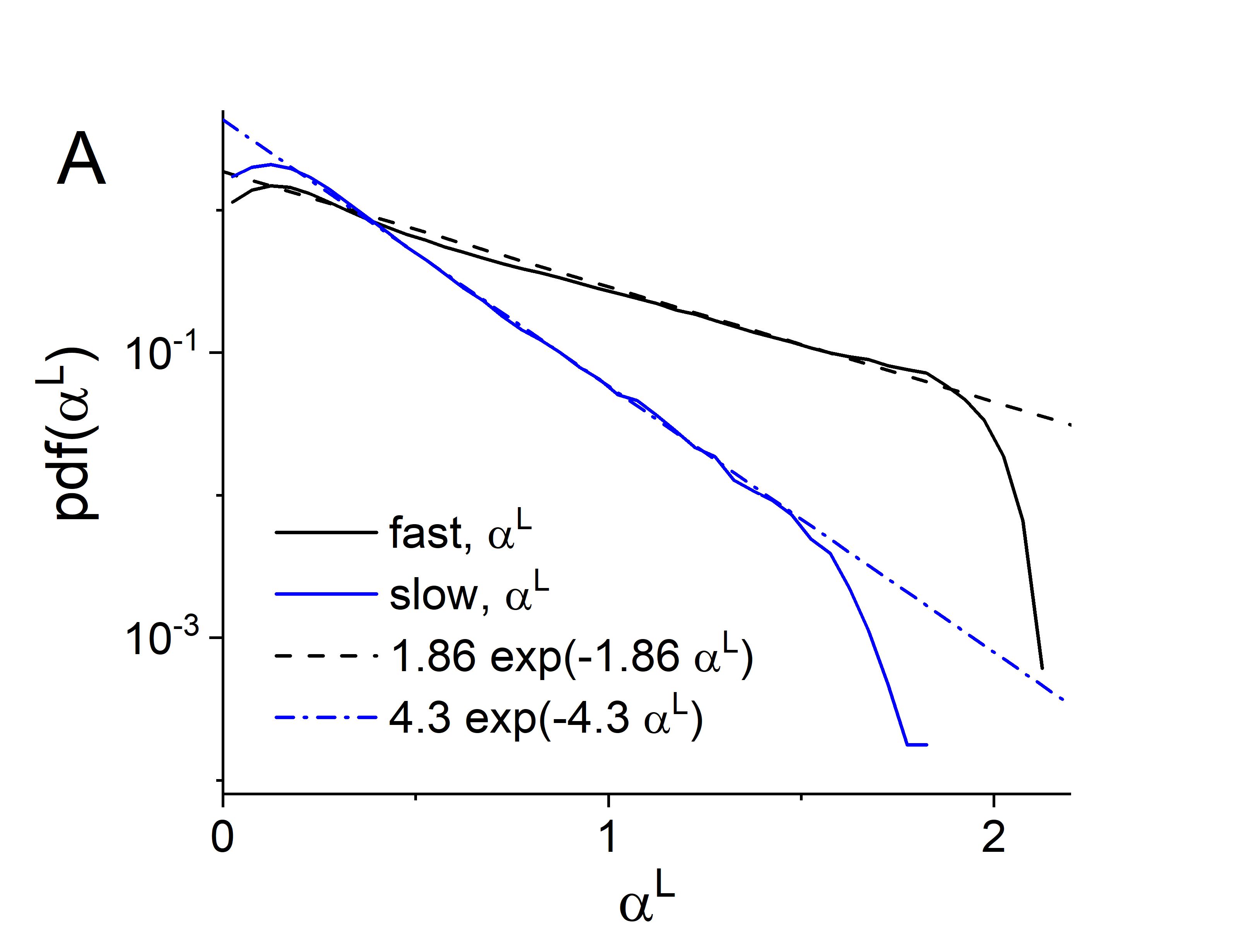}
\includegraphics[scale=0.3]{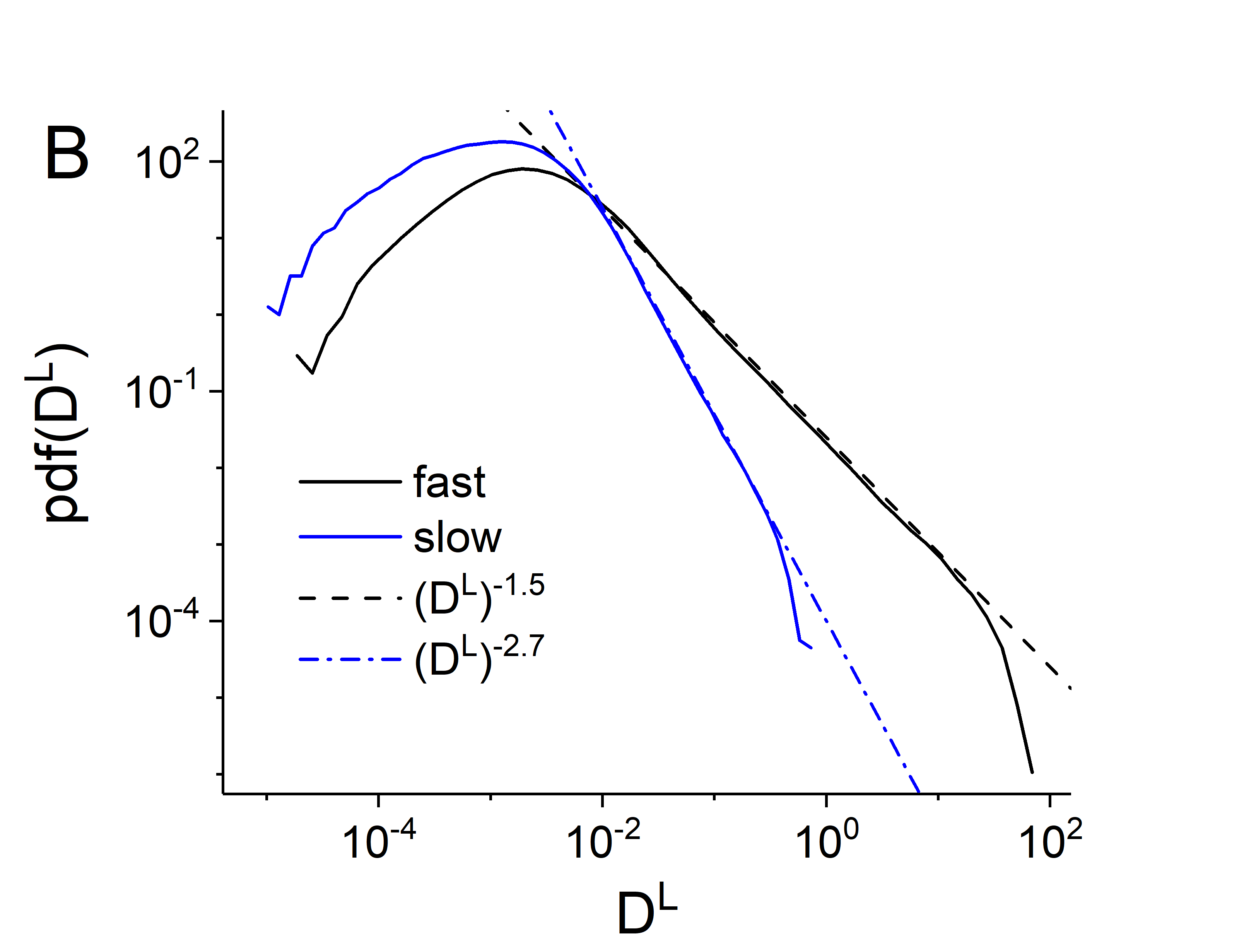}
\caption{Distribution of local anomalous exponents $\alpha^L$ (\textbf{A}) and local generalized diffusion coefficients $D^L$ (\textbf{B}) obtained from experimental trajectories of slow and fast endosomes. The dashed and dashed-dotted lines are best fits to exponential (\textbf{A}) and power-law pdfs (\textbf{B}). In (\textbf{A}) they correspond to $1.86 \exp(-1.86 \alpha^L)$ for pdf of $\alpha^L$ of fast endosomes (dashed line) and $4.3 \exp(-4.3 \alpha^L)$ for pdf of $\alpha^L$ of slow endosomes (dashed-dotted line). In (\textbf{B}) they correspond to $(D^L)^{-1.5}$ for pdf of $D^L$ of fast endosomes (dashed line) and $(D^L)^{-2.7}$ for pdf of $D^L$ of slow endosomes (dashed-dotted line). 
\label{fig2}}
\end{figure} 
The PDFs of $\alpha^L$ and $D_{\alpha^L}$ for slow and fast endosomes are shown in Figure~\ref{fig2} A and B. In both cases, the   PDFs of $\alpha^L$ follow an exponential distribution, while those of  $D_{\alpha^L}$  are best fitted with a power-law. However, the parameters characterizing the distribution shapes are very different. Furthermore the parameters for the fast endosomes PDFs coincide with those found by considering  all experimental trajectories \cite{N}. This is in agreement with an heterogeneous FBM model of endosomal transport \cite{N}, which describes the endosome motion as FBM with non-constant Hurst exponents.

\begin{figure}[htb]
\includegraphics[scale=0.3]{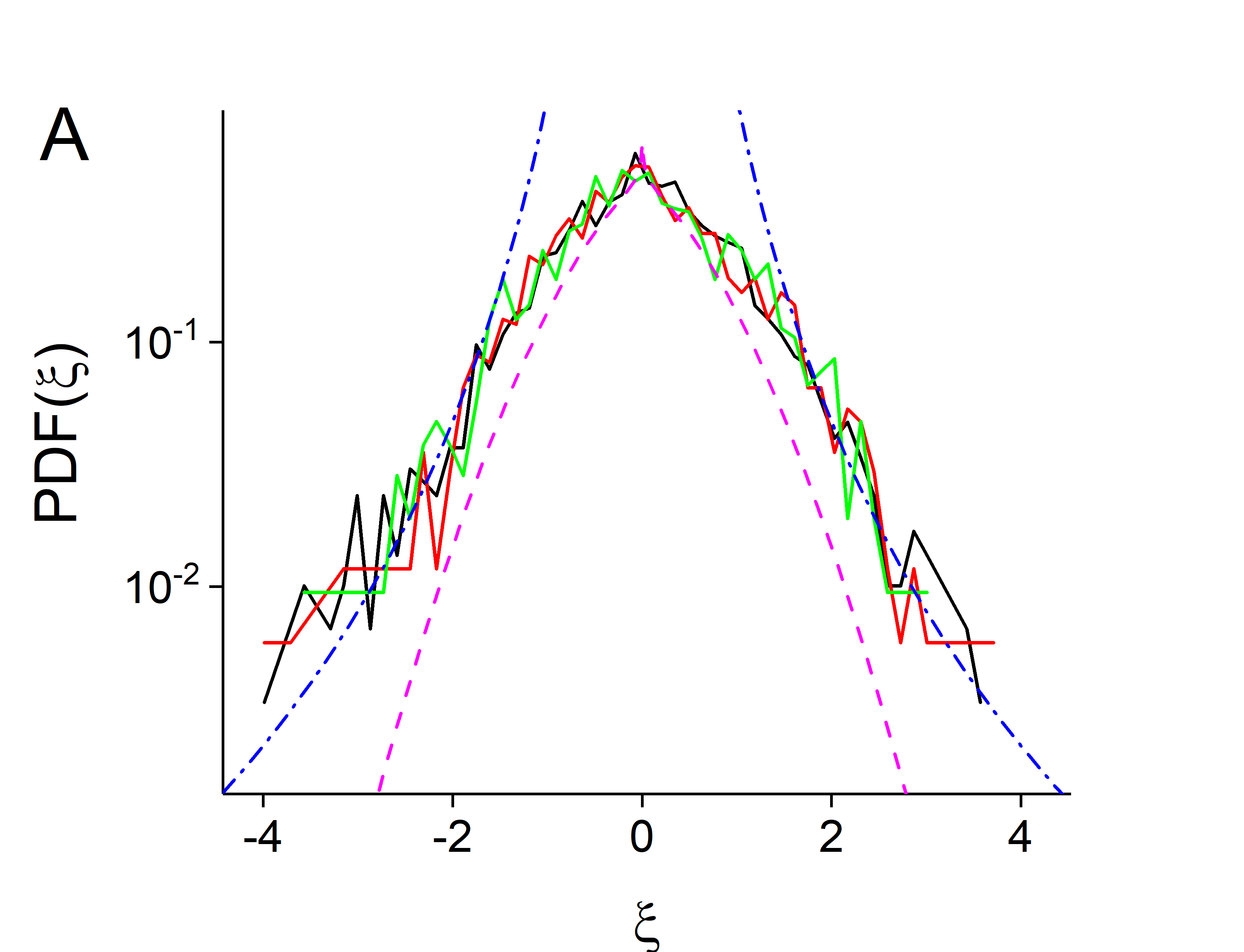}
\includegraphics[scale=0.3]{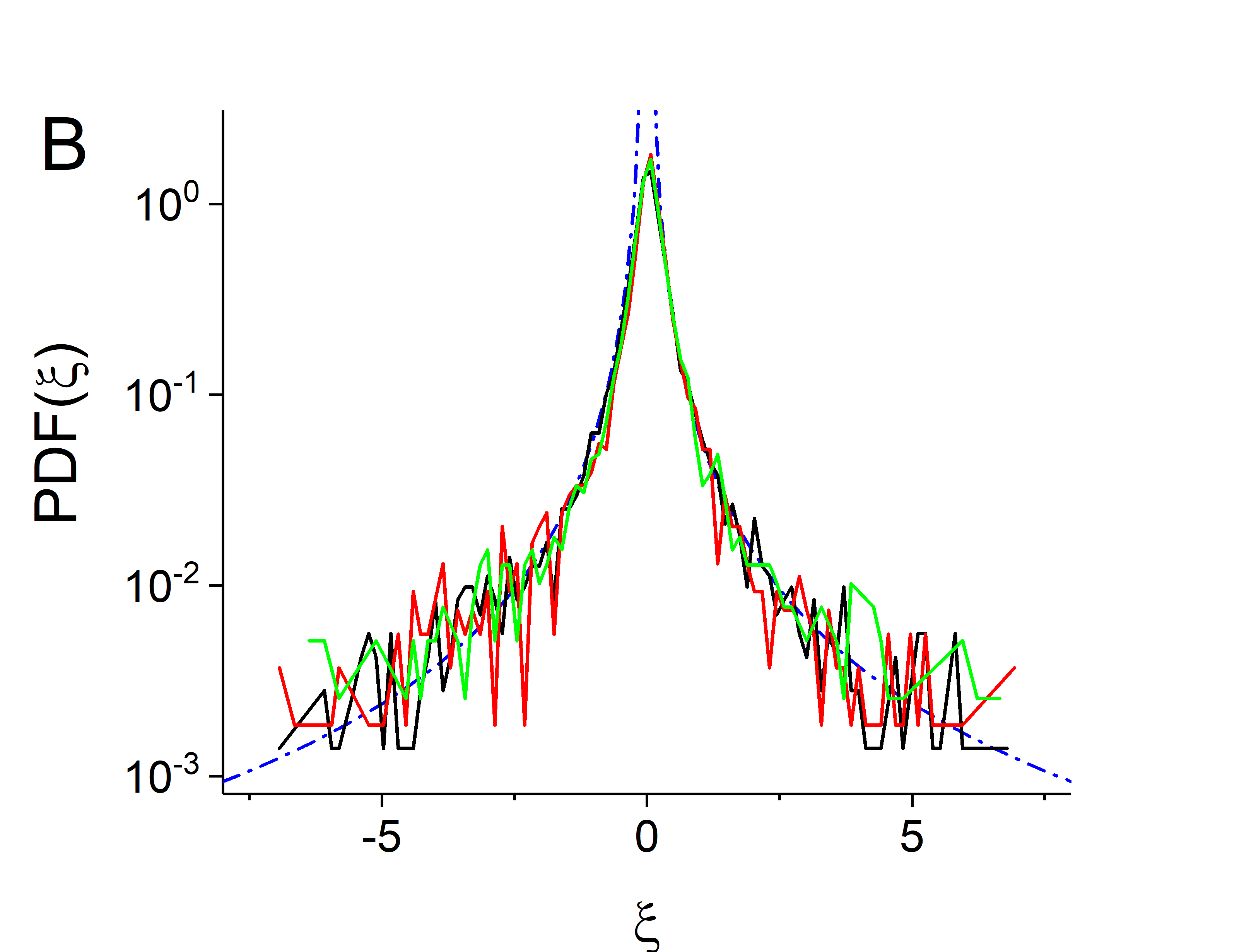}
\caption{
Distribution of scaled x-component of coordinate $\xi=x/\sigma_x$ obtained from experimental trajectories of slow (\textbf{A}) and fast (\textbf{B}) endosomes. The dashed-dotted lines correspond to power-law fit $|\xi|^{-1-2\gamma_S}$ for slow endosomes ($\gamma_S \simeq 1.7$) and $|\xi|^{-1-2\gamma_F}$ for fast endosomes ($\gamma_F \simeq 0.5$). In (\textbf{A}) we also compare PDF of slow endosomes with the analytical propagator for obstructed diffusion (dashed line) \cite{Krapf2012}. 
\label{figPROPAGATORS}}
\end{figure}

Finally, we calculated propagators of experimental trajectories for slow and fast endosomes (Fig. \ref{figPROPAGATORS}). Using the power-law forms of distributions of local generalized diffusion coefficients of slow $p_S(D^L) \sim (D^L)^{-1-\gamma_S}$ and fast $p_F(D^L) \sim (D^L)^{-1-\gamma_F}$ endosomes with $\gamma_S \simeq 1.7$ and $\gamma_F \simeq 0.5$ (Fig. \ref{fig2}), we fit the propagators with the propagators of hFBM, PDF$(\xi) \sim |\xi|^{-1-2\gamma}$ with $\gamma=\gamma_S$ for slow endosomes and $\gamma=\gamma_F$ for fast endosomes (see Supplementary Note and Ref. \cite{Chechkin}). For slow endosomes (Fig. \ref{figPROPAGATORS}A), we also compare the experimental PDFs with the analytical propagator for obstructed diffusion in two dimensions, $\xi^{-0.108} \exp (-|\xi|^{1.65})$ \cite{Krapf2012}. 

\section{Discussion}

In this paper we have extended our investigation of the heterogeneous intracellular transport of endosomes based on the local analysis of experimental trajectories \cite{N}. 
Individual endosomes move for long distances in a heterogeneous way with short bursts of directed motility, interspersed with periods of sub-diffusive motion \cite{Rodriguez,Zajac}. The heterogeneous character of this motion is also manifested as some endosomes are less motile then others. Some endosomes look as of they are jiggling in one position for the whole period of observation. Therefore, we split the ensemble of trajectories into slow and fast moving endosomes. The distinct time behaviour of mean squared displacements and velocity auto-correlation functions confirm the effectiveness of these methods. The splitting allowed us to study passive sub-diffusive and active super-diffusive transport of endosomes separately. 

Comparing the behaviour of fast endosomes (MSDs, VACFs and propagators) to the behaviour of the entire ensemble, we find that they are most consistent with FBM models \cite{N}. Therefore, we conclude that fast endosomes follow heterogeneous FBM \cite{N}. The ergodicity (Figure~\ref{fig1A}A) and the VACF  (Figure~\ref{fig3}A) suggest that slow endosomes are also described by the hFBM or heterogeneous generalized fractional Langevin equation motion. 
For slow endosomes, crowding and obstruction effects could also lead to  sub-diffusive behaviour \cite{HF,BGM}. 
It is known that obstructed diffusion has many similarities with fBm such as stationarity of the increments and the equivalence of the time and ensemble MSDs \cite{Krapf2015, Krapf2012}. The propagators provide a clear way to distinguish obstructed diffusion from fBm. Therefore, we calculated propagators of experimental slow endosomes and compare it with analytical prediction for the propagator of obstructed diffusion and prediction of heterogeneous fBM. The results shown in Fig. (\ref{figPROPAGATORS}) indicate that slow endosomes follow hfBm at longer time scales while on smaller scales obstructed diffusion likely contributes to their sub-diffusive behaviour as well. Crowding effects remain as a possible source of anomalous diffusion of slow endosomes. Recently, numerical simulations of lipids in crowded conditions of the membrane was shown to be multifractal and anomalous. The dynamics was no longer described by the mechanism consistent with the fractional Langevin equation, or by any single known mechanism. Instead, the motion was found to be non-Gaussian and heterogeneous, yet maintains its ergodic properties \cite{Jeon} which is similar to what we observed for experimental trajectories of slow endosomes.

Both slow and fast endosomal trajectories are found to be highly heterogeneous in space and time. The spatial heterogeneity in the form of coupling between endosome diffusivity and duration of endosome trajectory explains the behaviour of the MSDs. Longer trajectories have smaller generalized diffusion coefficients since in experiments slowly moving endosomes with smaller diffusion coefficients stay longer in the field of view, having longer durations. For slow and fast endosomes, we can conclude that EMSD and E-TMSD are not adequate to describe the large heterogeneity exhibited in space and time. Therefore, we applied a time local analysis of individual trajectories. 

From the local analysis, we found that slow and fast endosomal trajectories are both characterized by exponentially distributed anomalous exponents and power-law distributed generalized diffusion coefficients. However, the parameters of these distributions are different. Although the factors that cause the power-law distributed generalized diffusion coefficients for slow and fast endosomes could be different, some common factors can exist. One of them could be the scale free properties of endosomal networks \cite{Foret}. Hence the differences in endosome diameters could generate distinct diffusive properties intrinsic to each endosome. Heterogeneous diffusion generated by the fluctuations of molecular size was found in single-molecule experiments within the cell \cite{Lampo,SadoonWang,He}. Another common factor promoting power-law distributions of generalized diffusion coefficients could be non-specific interactions with the endoplasmic reticulum or other organelles and large intracellular structures. Recently, non-specific interactions were shown to generate heterogeneous diffusion of nanosized objects in mammalian cells \cite{Etoc}.

Our analysis of endosomal transport would be valuable both for fundamental cell biology and nanomedicine applications such as drug and gene delivery. Often in these applications, nanoparticles are used as cargo-carrying vesicles which in turn utilize endosomal network for their intracellular transport. For example, gold nanoparticles was shown to cluster inside endosomes and move via sub- and superdiffusion \cite{Liu}. Our results would be also useful for the nanoparticle enhanced radiation therapy of cancer \cite{Lin,Currell,Sotiropoulos} where clusters of nanoparticles inside endosomes are used for dose-enhancement.

In the future, we expect microscopy techniques will improve in tandem with tracking algorithms, providing data sets with larger ranges of time scales and improved resolution. Thus further subclassification of ensembles of endosomal tracks (beyond the binary fast and slow separation) will become possible towards the ultimate goal of single molecule specificity. Increasing the dynamic range (to sub millisecond time scales) will allow the stepping motion of the motor proteins (kinesin and dynein) attached to microtubules to be connected with the spectra of $\alpha$ and $D_{\alpha}$ for the fast moving endosomes at a fundamental level.


\begin{acknowledgments}
The authors thank: NK, SF and VA acknowledge financial support from EPSRC Grant No. EP/V008641/1. DH acknowledges financial support from the Wellcome Trust Grant No. 215189/Z/19/Z. GP is supported by the Basque Government through the BERC 2018-2021 programs and by the Spanish Ministry of Economy and Competitiveness MINECO through the BCAM Severo Ochoa excellence accreditation SEV-2017-0718.
\end{acknowledgments}

\appendix

\setcounter{figure}{0}
\makeatletter 
\renewcommand{\thefigure}{S\@arabic\c@figure}
\makeatother

\section*{Appendix A}

Figure A1: A example of experimental endosome trajectories measured in MRC5 cells stably expressing GFP-Rab5. See the main text for details.
\begin{figure}[htb]
\includegraphics[scale=0.7]{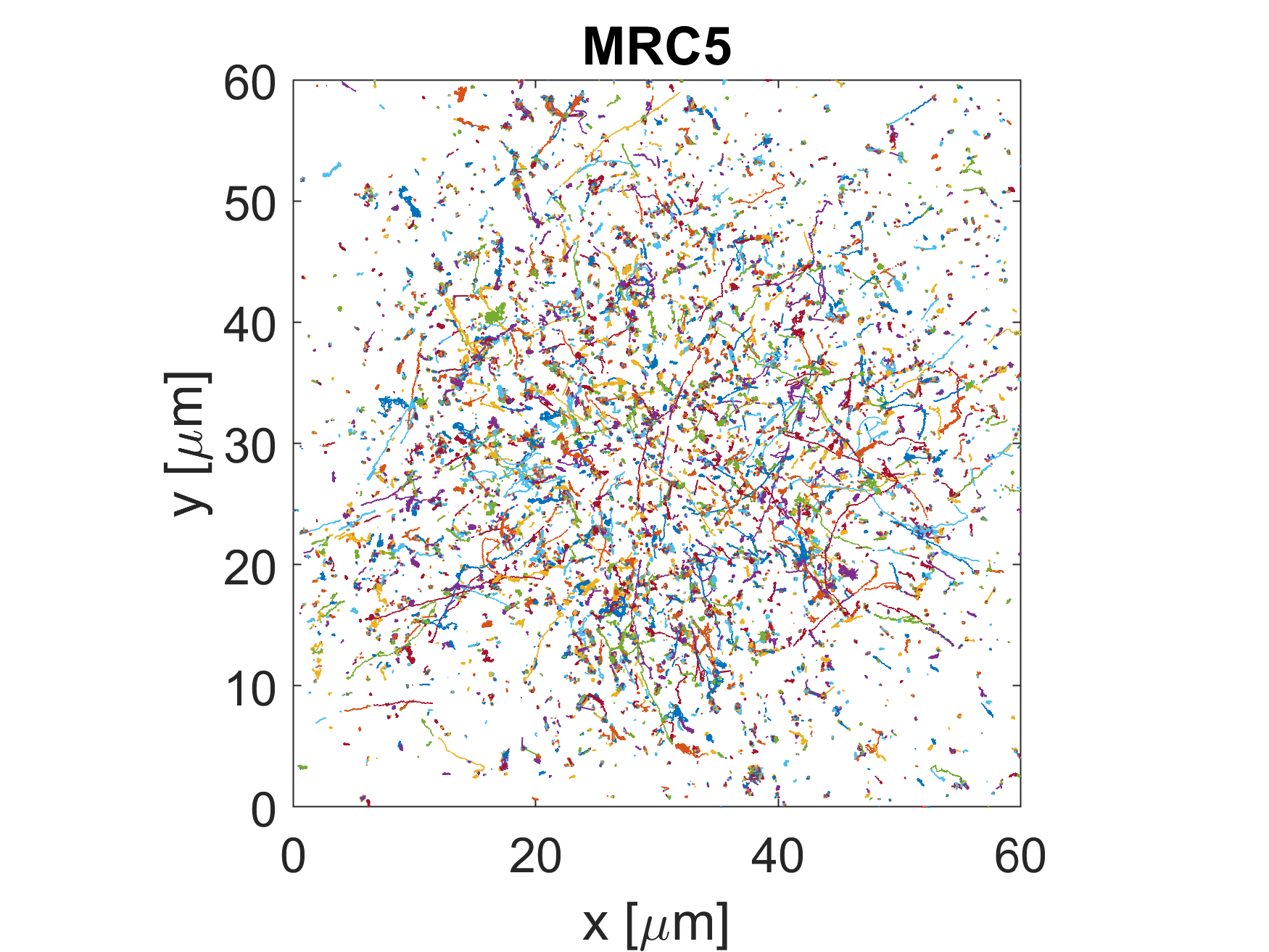}
\caption{An example of experimental endosomal trajectories ($30000$ trajectories are shown). 
\label{figS0}}
\end{figure}

Figure A2: Two splitting methods used to separate endosome trajectories into slow and fast moving endosomes. The first method uses the maximum distance traveled $R(t)$. The second method uses the time-dependent anomalous exponent $H(t)$ estimated with the neural network. An example of two trajectories is shown which were classified as slow and fast by both methods. See the main text for details of the methods.
\begin{figure}[htb]
\includegraphics[scale=0.3]{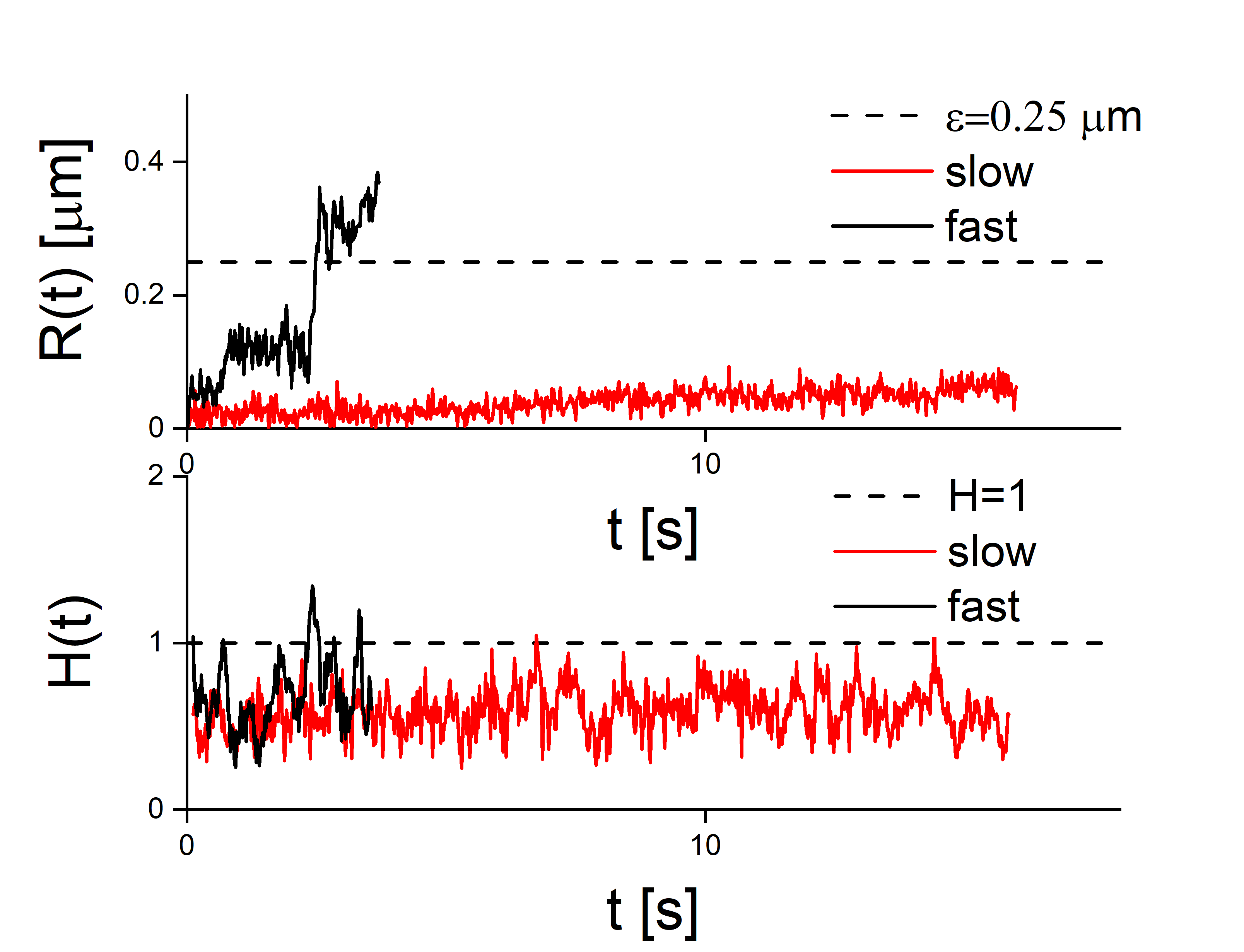}
\caption{An example of experimental trajectories of slow and fast moving endosomes obtained using the maximum distance traveled $R(t)$ (upper panel) and the time-dependent anomalous exponent $H(t)$ estimated with the neural network (lower panel).
\label{figS1}}
\end{figure}

Figure A3: The EMSD of slow and fast moving endosomes calculated with the splitting method which uses the maximum distance traveled $R(t)$. Two values of the threshold $\epsilon$ produce qualitatively similar results which suggests that the splitting method is robust against small variations of the threshold.
\begin{figure}[htb]
\includegraphics[scale=0.3]{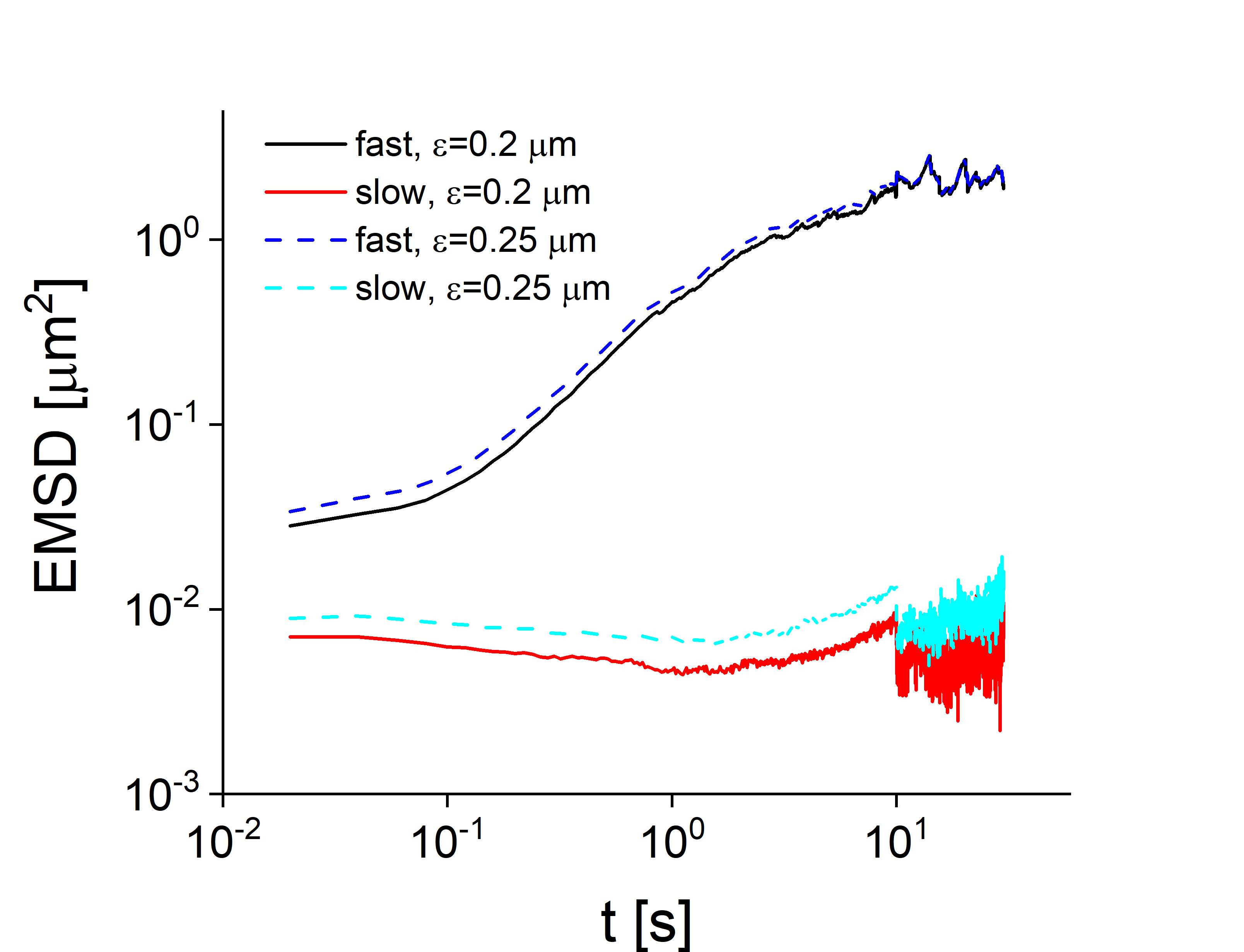}
\caption{The EMSD of experimental trajectories of slow and fast moving endosomes calculated with the splitting method which uses the maximum distance traveled $R(t)$ for two values of the threshold $\epsilon=0.2$ $\mu$m and $\epsilon=0.25$ $\mu$m. 
\label{figS1B}}
\end{figure}

Figure A4: Comparison of distributions of anomalous exponents $\alpha^{NN}$ and generalized diffusion coefficients $D^{NN}$ and local anomalous exponents $\alpha^L$ and $D^L$ of slow moving endosomes. Anomalous exponents $\alpha^{NN}$ were estimated using a neural network with window size $0.26$ s. The generalized diffusion coefficients $D^{NN}$ were estimated by fitting the local TMSD of the trajectory with the power law $D^{NN} t^{\alpha^{NN}}$. The distribution of $\alpha^{NN}$ has a maximum of $0.6$ and decays faster then the distribution of $\alpha^L$. This may be because many short trajectories are missing in the NN analysis, since the NN could analyse trajectories with durations longer than its window size \cite{ELife}. The distributions of generalized diffusion coefficients (right panel), on the other hand, are similar to each other.
\begin{figure}[htb]
\includegraphics[scale=0.3]{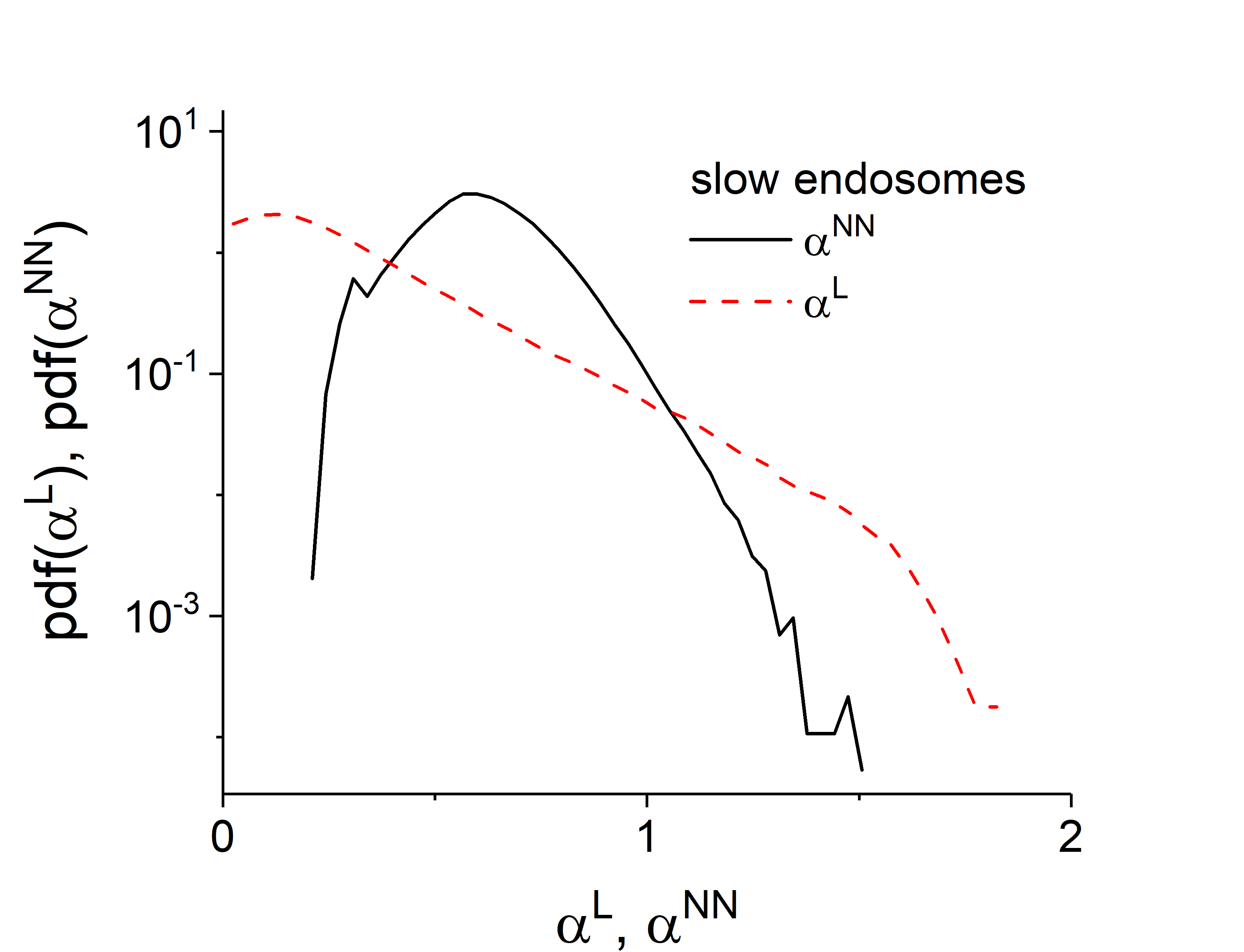}
\includegraphics[scale=0.3]{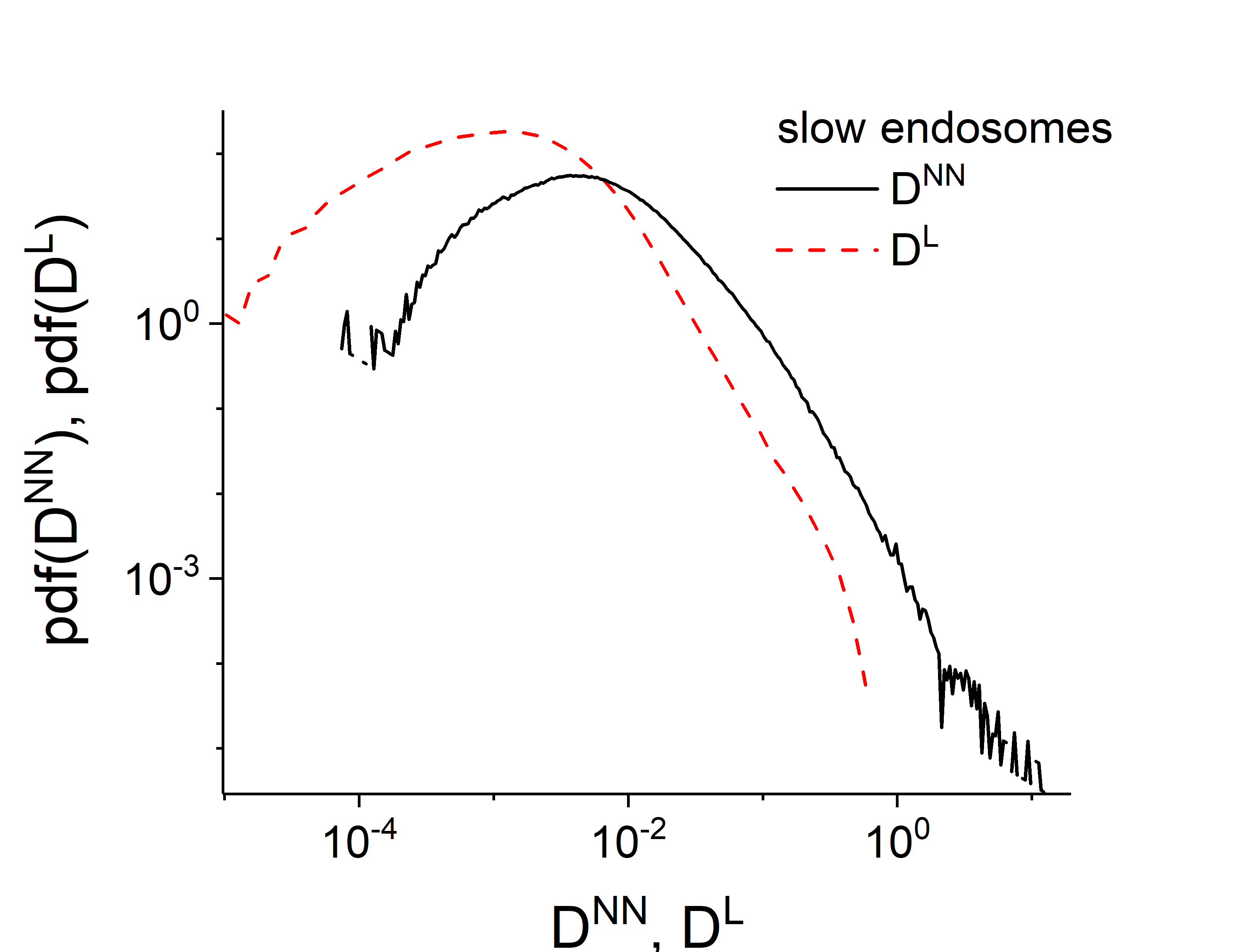}
\caption{Slow endosomes. Left panel: Distribution of anomalous exponents $\alpha^{NN}$ of slow moving endosome trajectories (the solid curve) compared with the distribution of local anomalous exponents $\alpha^L$ of slow moving endosome trajectories (the dashed curve). Right panel: Distribution of generalized diffusion coefficients $D^{NN}$ of slow moving endosome trajectories (the solid curve) compared with the distribution of local generalized diffusion coefficients $D^L$ of slow moving endosome trajectories (the dashed curve).
\label{figS2}}
\end{figure} 

Figure A5: Comparison of distributions of anomalous exponents $\alpha^{NN}$ and generalized diffusion coefficients $D^{NN}$ and local anomalous exponents $\alpha^L$ and $D^L$ of fast moving endosomes. Anomalous exponents $\alpha^NN$ were estimated using neural network with window size $0.26$ s. The generalized diffusion coefficients $D^{NN}$ were estimated by fitting the local TMSD of trajectory with the power law $D^{NN} t^{\alpha^{NN}}$. The distributions of anomalous exponents (left panel) are similar to each other while the distributions of generalized diffusion coefficients (right panel) are almost indistinguishable. 
\begin{figure}[htb]
\includegraphics[scale=0.3]{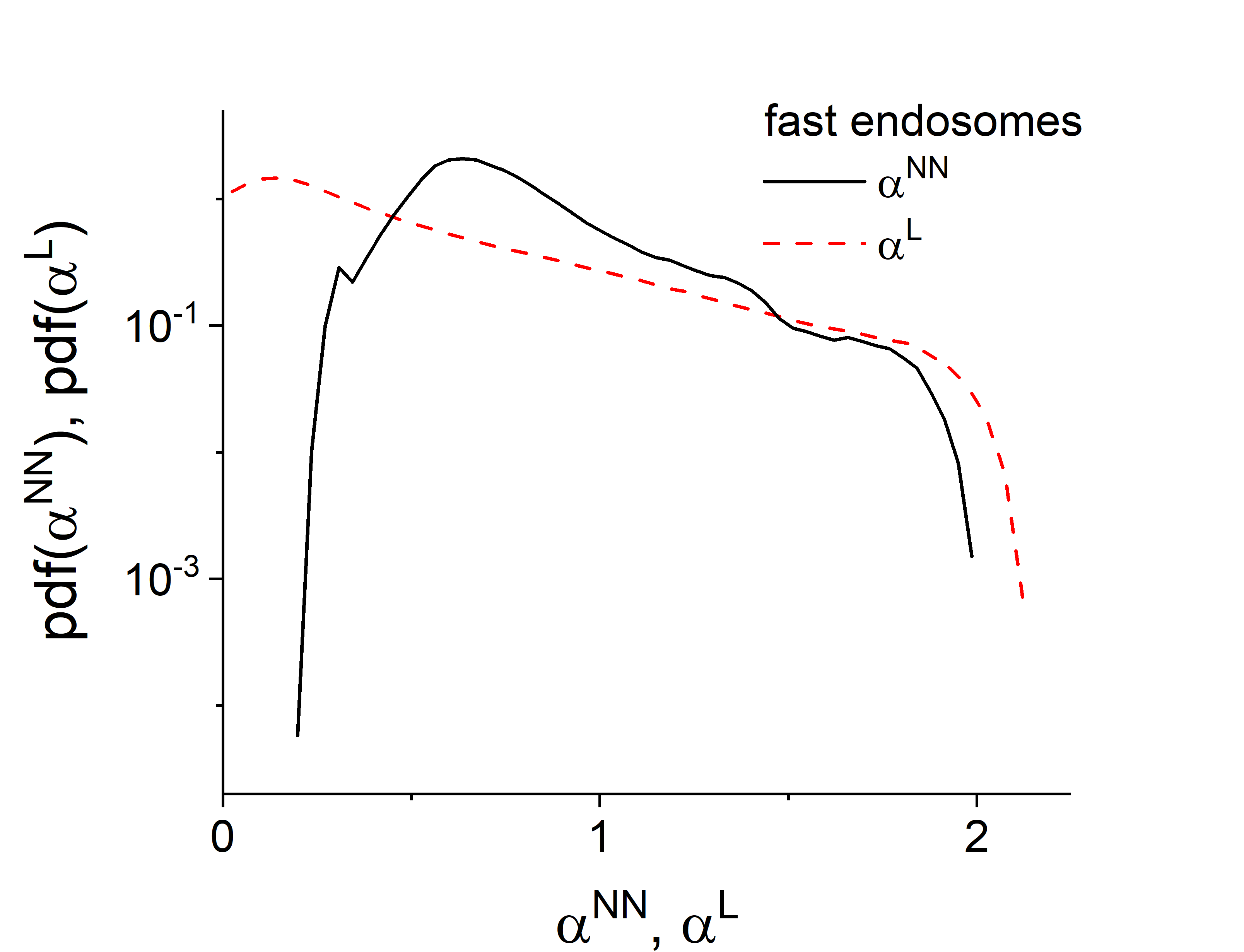}
\includegraphics[scale=0.3]{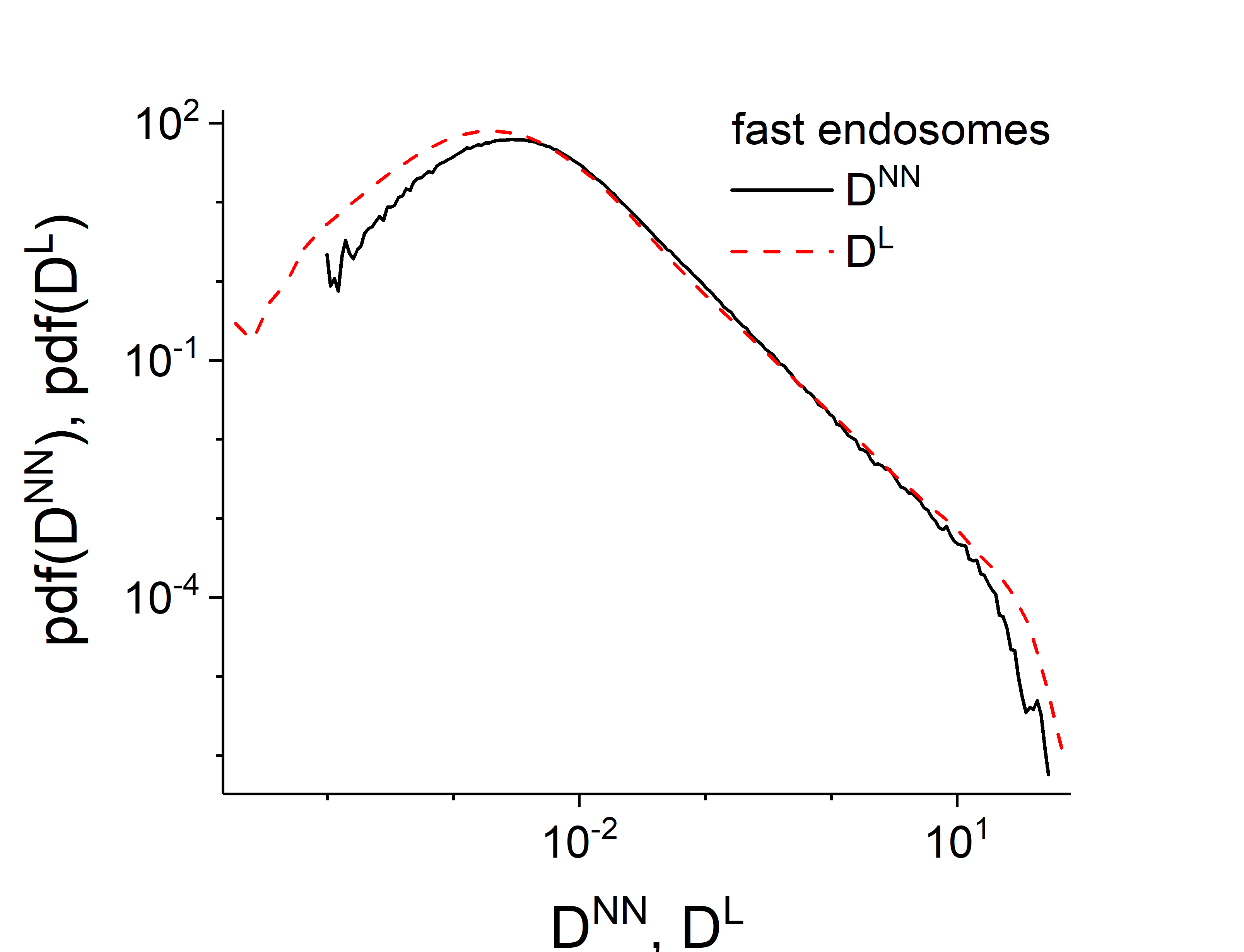}
\caption{Fast endosomes. Left panel: Distribution of anomalous exponents $\alpha^{NN}$ of fast moving endosome trajectories (the solid curve) compared with the distribution of local anomalous exponents $\alpha^L$ of slow moving endosome trajectories (the dashed curve). Right panel: Distribution of generalized diffusion coefficients $D^{NN}$ of fast moving endosome trajectories (the solid curve) compared with the distribution of local generalized diffusion coefficients $D^L$ of fast moving endosome trajectories (the dashed curve).
\label{figS3}}
\end{figure} 

Figure A6: Local anomalous exponents $\alpha^{L}$ and local generalized diffusion coefficients $D^{L}$ are positively correlated both for slow and fast moving endosomes. 
\begin{figure}[htb]
\includegraphics[scale=0.7]{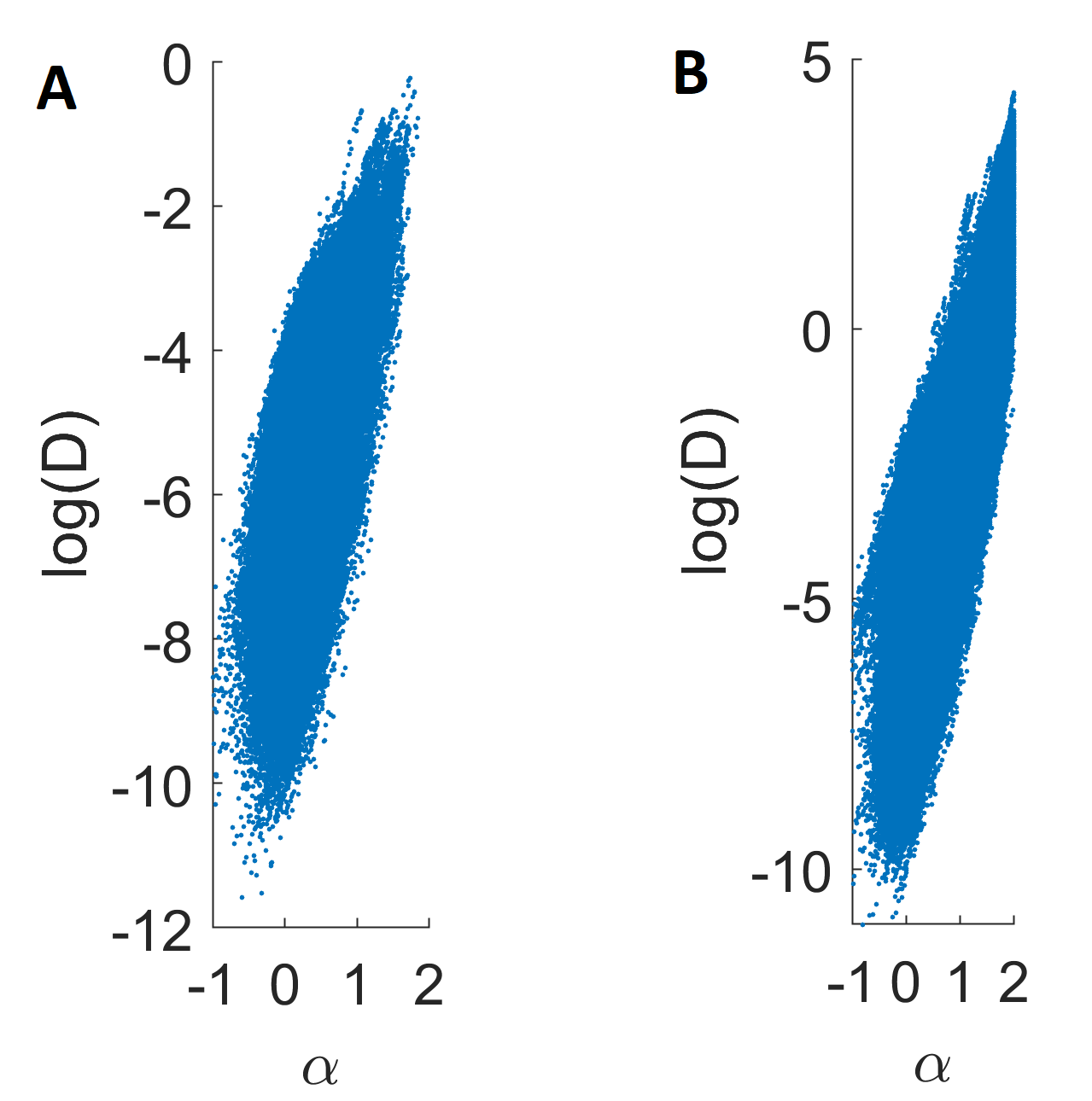}
\caption{
 Correlation between local anomalous exponents $\alpha^{L}$ and local generalized diffusion coefficients $D^{L}$ for slow (A) and fast (B) moving endosomes.
\label{figS5}}
\end{figure}

\setcounter{equation}{0}


\begin{thebibliography}{999}

\bibitem{Klages} Klages, R., Radons, G., and Sokolov, I.M. (eds.). 2004. Anomalous Transport: Foundations and Applications. Wiley VCH - Verlag, Weinheim.

\bibitem{HF} Hofling, F., and Franosch, T. Anomalous transport in the crowded world of biological cells, {\em Rep. Prog. Phys.} {\bf 2013} {\em 76}, 046602.


\bibitem{Metzler2014} Metzler, R., Jeon, J.-H., Cherstvy, A. G., and Barkai, E. Anomalous diffusion models and their properties: non-stationarity, non-ergodicity, and ageing at the centenary of single particle tracking. {\em Phys. Chem. Chem. Phys.} {\bf 2014} {\em 16}, 24128.

\bibitem{BGM}
Barkai, E., Garini, Y., and Metzler, R.\ Strange kinetics of single molecules in living cells. {\em Phys. Today} {\bf 2012}, {\em 65}, 29-35.

\bibitem{Heinrich} Arcizet, D.\, Meier, B.\, Sackmann, E.\, Radler, J.\ O.\ and Heinrich, D.\  Temporal Analysis of Active and Passive Transport in Living Cells. {\em Phys. Rev. Lett.} {\bf 2008}, {\em 101}, 248103. 

\bibitem{WaighBOOK} Waigh, T.A. 2014. The Physics of Living Processes: A Mesoscopic Approach. John Wiley \& Sons.

\bibitem{Akimoto}
Akimoto, T. and Yamamoto, E.
Detection of transition times from single-particle-tracking trajectories.
{\em Phys. Rev. E} {\bf 2017}, {\em 96}, 052138.

\bibitem{ELife} 
Han, D., Korabel, N., Chen, R., Johnston, M., Gavrilova, A., Allan, V.J., Fedotov, S., Waigh, T. A. Deciphering anomalous heterogeneous intracellular transport with neural networks. {\em eLife} {\bf 2020}, {\em 9}, e52224.

\bibitem{Newby}  Newby, J. M., Schaefer, A. M., Lee, P. T., Forest, M. G., Lai, S. K. Convolutional neural networks automate detection for tracking of submicron-scale particles in 2D and 3D. {\em PNAS} {\bf 2018}, {\em 115}, 9026-9031.


\bibitem{Saxton} 
Saxton, M.\ J.\ 
Single-particle tracking: the distribution of diffusion coefficients. 
{\em Biophys.\ J.} {\bf 1997}, {\em 72}, 1744--1753.

\bibitem{Granick1} 
Wang, B., Anthony, S.\ M., Bae, S.\ C. and Granick, S. 
Anomalous yet Brownian. 
{\em PNAS} {\bf 2009}, {\em 106}, 15160. 

\bibitem{Granick2} 
Wang, B., Kuo, J., Bae, S.\ C., Granick, S.\ 
When Brownian diffusion is not Gaussian. 
{\em Nat. Mater.} {\bf 2012}, {\em 11}, 481--485.

\bibitem{Lampo} 
Lampo, T.\ J., Stylianidou, S.\, Backlund, M.\ P.\,  Wiggins, P.\ A.\ and Spakowitz, A.\ J.\ 
Cytoplasmic RNA-Protein Particles Exhibit Non-Gaussian Subdiffusive Behavior. 
{\em Biophys.\ J.} {\bf 2017}, {\em 112}, 532--542.

\bibitem{Weiss2020} 
Sabri, A., Xu, X., Krapf, D., and Weiss M. 
Elucidating the Origin of Heterogeneous Anomalous Diffusion in the cytoplasm of mammalian cells. 
{\em Phys. Rev. Lett.} {\bf 2020}, {\em 125}, 058101.

\bibitem{BaLysosomes} 
Ba, Q., Raghavan, G., Kiselyov, K. \& Yang, G. 
Whole-cell scale dynamic organization of lysosomes revealed by spatial statistical analysis. 
{\em Cell Reports} {\bf 2018}, {\em 23}, 3591--3606.

\bibitem{ManzoPRX} 
Manzo, C., Torreno-Pina, J.\ A., Massignan, P., Lapeyre, Jr., G.\ J., Lewenstein, M.\ and Garcia-Parajo, M.\ F.\ 
Weak Ergodicity Breaking of Receptor Motion in Living Cells Stemming from Random Diffusivity. 
{\em Phys.\ Rev.\ X} {\bf 2015}, {\em 5}, 011021.

\bibitem{SadoonWang} 
Sadoon, A.\ A.\  and Wang, Y.\ 
Anomalous, non-Gaussian, viscoelastic, and age-dependent dynamics of histonelike nucleoid-structuring proteins in live Escherichia coli. 
{\em Phys.\ Rev.\ E} {\bf 2018}, {\em 98}, 042411.

\bibitem{Calderon} Calderon, C. P. Motion blur filtering: A statistical approach for extracting confinement forces and diffusivity from a single blurred trajectory. {\em Phys. Rev. E} {\bf 2016}, {\em 93}, 053303.

\bibitem{Andersson} Godoy, B. I., Vickers, N. A. and Andersson, S. B. An Estimation Algorithm for General Linear Single Particle Tracking Models with Time-Varying Parameters. {\em Molecules} {\bf 2021}, {\em 26}, 886.

\bibitem{Holcman} HozÃ© N. and Holcman, D. Statistical Methods for Large Ensembles of Super-Resolution Stochastic Single Particle Trajectories in Cell Biology, 
{\em Annu. Rev. Stat. Appl.} {\bf 2017}, {\em 4}, 189-223.

\bibitem{Weron} Weron, A. Mathematical Models for Dynamics of Molecular Processes in Living Biological Cells. A Single Particle Tracking Approach, {\em Ann. Math. Sil} {\bf 2018} {\em 32}, 5-41.

\bibitem{Weiss2014} Ernst, D., KÃ¶hler, J., Weiss, M. Probing the type of anomalous diffusion with single-particle tracking. {\em Phys. Chem. Chem. Phys.} {\bf 2014}, {\em 16}, 7686-7691.


\bibitem{Janczura} J. Janczura, M. Balcerek, K. Burnecki, A. Sabri, M. Weiss and D. Krapf, Identifying heterogeneous diffusion states in the cytoplasm by a hidden Markov mode. {\em New J. Phys.} {\bf 2021}, {\em 23}, 053018.

\bibitem{Metzler2020} 
Metzler, R. Superstatistics and Non-Gaussian. {\em Eur.\ Phys.; J.\ Special Topics} {\bf 2020}, {\em 229}, 711--728 (2020).

\bibitem{Beck1} Beck, C. and Cohen, E. D. B. 
Superstatistics. 
{\em Physica A} {\bf 2003}, {\em 322}, 267.

\bibitem{Molina} 
Molina-Garc\'ia, D., Pham, T.M., Paradisi, P., Manzo, C. and Pagnini, G. 
Fractional kinetics emerging from ergodicity breaking in random media, 
{\em Phys. Rev. E} {\bf 2016}, {\em 94}, 052147.

\bibitem{Mackala} 
Maćkała, A. and Magdziarz, M. 
Statistical analysis of superstatistical fractional Brownian motion and applications. 
{\em Phys.\ Rev.\ E} {\bf 2019}, {\em 99}, 012143.

\bibitem{Chertvy2014} 
Cherstvy, A.\ G.\, and Metzler, R.\ 
Nonergodicity, fluctuations, and criticality in heterogeneous diffusion processes. 
{\em Phys.\ Rev. E} {\bf 2014}, {\em 90}, 012134.

\bibitem{Spakowitz} 
Spakowitz, A.\ J.\ 
Transient Anomalous Diffusion in a Heterogeneous Environment.
{\em Front. Phys.} {\bf 2019}, {\em 7}, 119. 

\bibitem{Chubynsky} 
Chubynsky, M.\ V.\ and Slater, G.\ W.\ 
Diffusing diffusivity: a model for anomalous yet Brownian diffusion. 
{\em Phys.\ Rev.\ Lett.} {\bf 2014}, {\em 113}, 098302.

\bibitem{Sposini} 
Sposini, V., Chechkin, A.V., Seno, F., Pagnini, G. and Metzler, R. 
Random diffusivity from stochastic equations: comparison of two models for Brownian yet non-Gaussian diffusion. 
{\em New J. Phys.} {\bf 2018}, {\em 20}, 043044.

\bibitem{Chechkin} 
Chechkin, A.V., Seno, F., Metzler, R. and Sokolov, I.M. 
Brownian yet Non-Gaussian Diffusion: From Superstatistics
to Subordination of Diffusing Diffusivities. 
{\em Phys. Rev. X} {\bf 2017}, {\em 7}, 021002. 

\bibitem{Wang} 
Wang, W., Cherstvy, A. G., Chechkin, A. V., Thapa, S., Seno, F., Liu, X., and Metzler, R. Fractional Brownian motion with random diffusivity: emerging residual nonergodicity below the correlation time. 
{\em J. Phys. A: Math. Theor.} {\bf 2020}, {\em 53}, 474001.

\bibitem{itto-beck-2021}
Itto, Y.\ and Beck, C.\
Superstatistical modelling of protein diffusion dynamics in bacteria. 
{\em J. R. Soc. Interface} {\bf 2021}, {\em 18}, 20200927.

\bibitem{KB2010} 
Korabel, N., Barkai, E. 
Paradoxes of subdiffusive infiltration in disordered systems. 
{\em Phys. Rev. Let.} {\bf 2010}, {\em 104}, 170603.

\bibitem{FK2015} 
Fedotov S. and Korabel, N. 
Self-organized anomalous aggregation of particles performing nonlinear and non-Markovian random walks. 
{\em Phys. Rev. E} {\bf 2015}, {\em 92}, 062127.

\bibitem{FH2019} 
Fedotov, S. and Han, D. 
Asymptotic behavior of the solution of the space dependent variable order fractional diffusion equation: ultraslow anomalous aggregation. 
{\em Phys.\ Rev.\ Lett.} {\bf 2019}, {\bf 123}, 050602.

\bibitem{Sandev} 
Sandev, T., Chechkin, A. V., Korabel, N., Kantz, H., Sokolov, I. M., Metzler, R. 
Distributed-order diffusion equations and multifractality: Models and solutions. 
{\em Phys. Rev. E} {\bf 2015}, {\em 92}, 042117.

\bibitem{N}
Korabel, N., Han, D., Taloni, A., Pagnini, G., Fedotov, S., Allan, V., and Waigh, T. A., arXiv:2107.07760 [q-bio.SC].


\bibitem{Rogers} Rogers, S. S. Flores-Rodriguez, N., Allan, V. J., Woodman, P. G. and Waigh, T. A. The first passage probability of intracellular particle trafficking. 
{\em  Phys. Chem. Chem. Phys.} {\bf 2010}, {\em 12}, 3753-3761.

\bibitem{He} He, W., Song, H., Su, Y., Geng, L., Ackerson, B. J., Peng, H. B. and Tong, P. Dynamic heterogeneity and non-Gaussian statistics for acetylcholine receptors on live cell membrane. {\em Nat. Commun.} {\bf 2016}, {\em 7}, 11701.

\bibitem{Weber} Weber, S.\ C.\, Thompson, M.\ A.\, Moerner, W.\ E.\, Spakowitz, A.\ J.\ and Theriot, J.\ A.\ Analytical Tools To Distinguish the Effects of Localization Error, Confinement, and Medium Elasticity on the Velocity Autocorrelation Function. {\em Biophys.\ J.} {\bf 2012}, {\em 102}, 2443--2450.

\bibitem{Etoc} Etoc, F. {\it et al.} Non-specific interactions govern cytosolic diffusion of nanosized objects in mammalian cells. {\em Nat. Mater.} {\bf 2018}, {\em 17} 740--746.

\bibitem{Weber2010} Weber, S. C., Spakowitz, A. J. and Theriot, J. A. Bacterial Chromosomal Loci Move Subdiffusively through a Viscoelastic Cytoplasm. {\em Phys. Rev. Lett.} {\bf 2010}, {\em 104}, 238102.

\bibitem{WaighReview} Waigh, T.A. Advances in the microrheology of complex fluids, {\em Rep. Prog. Phys.} {\bf 2016}, {\em 79}, 74601-74663.

\bibitem{Doyle} Savin T. and Doyle P. S. Static and Dynamic Errors in Particle Tracking Microrheology. {\em Biophys. J.} {\bf 88}, {\em 2005}, 623-638. 


\bibitem{Rodriguez} Flores-Rodriguez, N., Rogers, S.S., Kenwright, D.A., Waigh, T.A., Woodman, P.G., Allan, V.J. Roles of dynein and dynactin in early endosome dynamics revealed using automated tracking and global analysis. {\em PLoS ONE} {\bf 2011}, {\em 6}, e24479.

\bibitem{Zajac} Zajac, A.L., Goldman, Y.E., Holzbaur, E.L., Ostap, E.M. Local cytoskeletal and organelle interactions impact molecular motor-driven early endosomal trafficking. {\em Curr. Biol.} {\bf 2013}, {\em 23}, 1173--80.

\bibitem{Krapf2015} Krapf, D. Mechanisms underlying anomalous diffusion in the plasma membrane. {\em Curr Top Membr.} {\bf 2015}, {\em 75}, 167-207.

\bibitem{Krapf2012} Weigel, A.V., Ragi, S., Reid, M.L., Chong, E.K.P., Tamkun, M.M. and Krapf, D. Obstructed diffusion propagator analysis for single-particle tracking. {\em Phys. Rev. E} {\bf 2012}, {\em 85}, 041924.

\bibitem{Jeon} Jeon, J. H., Javanainen, M., Martinez-Seara, H., Metzler, R. and Vattulainen, I., Protein Crowding in Lipid Bilayers Gives Rise to Non-Gaussian Anomalous Lateral Diffusion of Phospholipids and Proteins. {\em Phys. Rev. X} {\bf 2016}, {\em 6}, 021006.


\bibitem{Foret} Foret, L., Dawson J.E., VillaseÃ±or R., Collinet C., Deutsch A., Brusch L., Zerial M., Kalaidzidis Y., JÃ¼licher F. A general theoretical framework to infer endosomal network dynamics from quantitative image analysis. {\em Curr. Biol.} {\bf 2012}, {\em 22} 1381--1390. 

\bibitem{Liu} Liu, M., Li, Q., Liang, L. {\it et al.} Real-time visualization of clustering and intracellular transport of gold nanoparticles by correlative imaging. {\em Nat Commun.} {\bf 2017}, {\em 8}, 15646.

\bibitem{Lin} Lin, Y., McMahon,  S. J., Paganetti,  H., Schuemann, J. Biological modeling of gold nanoparticle enhanced radiotherapy for proton therapy. {\em Phys. Med. Biol.} {\bf 2015}, {\em 60}, 4149.

\bibitem{Currell} Villagomez-Bernabe, B. and Currell, F. J. Physical Radiation Enhancement Effects Around Clinically Relevant Clusters of Nanoagents in Biological Systems. {\em Sci. Rep.} {\bf 2019}, {\em 9}, 8156.  

\bibitem{Sotiropoulos} Sotiropoulos, M., Henthorn, N. T., Warmenhoven, J. W., Mackay, R. I., Kirkby, K. J., Merchant, M. J. Modelling direct DNA damage for gold nanoparticle enhanced proton therapy. {\em Nanoscale} {\bf 2017}, {\em 9}, 18413.


\end{thebibliography}

\end{document}